\documentclass[]{pasj01}
\draft
\usepackage{natbib}
\usepackage{lscape}

\begin{document} 
\Received{}
\Accepted{}

\title{AKARI mid-infrared slit-less spectroscopic catalogue}

\author{Mitsuyoshi \textsc{Yamagishi}\altaffilmark{1}}%
\altaffiltext{1}{Institute of Space and Astronautical Science, Japan Aerospace Exploration Agency, 3-1-1 Yoshinodai, Chuo-ku, Sagamihara, Kanagawa 252-5210, Japan}
\email{yamagish@ir.isas.jaxa.jp}
\author{Issei \textsc{Yamamura},\altaffilmark{1,2}}
\altaffiltext{2}{Department of Space and Astronautical Science, SOKENDAI, 3-1-1 Yoshinodai, Chuo-ku, Sagamihara, Kanagawa 252-5210, Japan}
\author{Toshiyuki \textsc{Mizuki}\altaffilmark{1}}
\author{Takafumi \textsc{Ootsubo}\altaffilmark{1}}
\author{Shunsuke \textsc{Baba}\altaffilmark{1}}
\author{Fumihiko \textsc{Usui}\altaffilmark{3}}
\altaffiltext{3}{Center for Planetary Science, Graduate School of Science, Kobe University, 7-1-48, Minatojima-Minamimachi, Chuo-Ku, Kobe 650-0047, Japan}
\author{Takashi \textsc{Onaka}\altaffilmark{4}}
\altaffiltext{4}{Graduate School of Science, The University of Tokyo, Bunkyo-ku, 113-0033, Tokyo, Japan}



\maketitle

\begin{abstract}
AKARI/IRC has a capability of the slit-less spectroscopy in the mid-infrared (5--13~$\micron$) over a 10$\arcmin$ $\times$ 10$\arcmin$ area with a spectral resolution of 50, which is suitable for serendipitous surveys.
The data reduction is, however, rather complicated by the confusion of nearby sources after dispersing the spectra.
To make efficient and reliable data reduction, we first compiled a point-source list from the reference image in each field-of-view and checked the overlaps of the spectra using their relative positions and fluxes.
Applying this procedure to 886 mid-infrared slit-less spectroscopic data taken in the cryogenic phase, we obtained 862 mid-infrared spectra from 604 individual non-overlapping sources brighter than 1.5~mJy.
We find a variety of objects in the spectroscopic catalogue, ranging from stars to galaxies.
We also obtained a by-product catalogue of 9~$\micron$ point sources containing 42,387 objects brighter than 0.3~mJy.
The spectroscopic and point-source catalogues are available online.
\end{abstract}

\section{Introduction}

The infrared camera (IRC) on-board the Japanese infrared astronomy satellite AKARI (\citealt{Murakami07, Onaka07}) has a slit-less spectroscopic capability in the near-infrared (NIR) and mid-infrared (MIR) ranges in addition to a slit spectroscopic capability (\citealt{Ohyama07}).
In the slit spectroscopic mode, AKARI/IRC observes the point sources and diffuse objects targeted by the observers.
Due to the simplicity of the observation and the data analysis, the slit spectroscopy is a solid technique to obtain spectra as used in Spitzer/IRS (\citealt{Houck04}).
Meanwhile, slit-less spectroscopy is generally used for surveying purposes.
Since all the light in the field-of-view (FOV) are dispersed at one time, slit-less spectroscopy is suitable for collecting unbiased spectral samples in serendipitous surveys.
The IRAS/LRS performed slit-less spectroscopy on bright point sources ($>$2~Jy) in the 8--23~$\micron$ range (\citealt{Olnon86}).
Slit-less spectroscopy with high sensitivity in the NIR and MIR is unique to AKARI/IRC until JWST/MIRI (\citealt{Rieke15, WrightGS15}) will be available.
NIR and MIR spectra show the emission features of polycyclic aromatic hydrocarbons, dust grains, and ices; therefore, the AKARI slit-less spectra are especially useful for probing the interstellar medium (ISM).

Slit-less spectroscopy, however, confuses the nearby sources after dispersion of the spectra, which complicates the data analysis.
If the FOV contains only a few bright sources, the overlap may be negligible, and we can extract their spectra without considering overlap.
In contrast, when the FOV contains many sources with similar fluxes, spectra of nearby sources are merged.
To avoid spectral contamination from nearby sources, we must select only the isolated point sources in the FOV.
For efficiency purpose, we use point-source information in the corresponding image for a source-position reference (the ``reference image'') taken by AKARI/IRC spectroscopic observations without a disperser.
From the optical design, we can infer how the point source is dispersed on the detector.
Therefore, based on the relative position and flux in the reference image, we can check whether or not the spectrum of an object overlaps with the spectra of nearby objects.

To extract as many uncontaminated spectra as possible, the small number of point sources in the FOV and/or the low spectral resolution are preferred.
Based on the latter idea, \citet{Shimonishi13} analyzed the NIR low-resolution slit-less data ($R\sim$20; 2--5~$\micron$) obtained as part of the mission program LSLMC (Large-area Survey of the Large Magellanic Cloud; PI: T. Onaka).
In the present paper, based on the former idea, we analyze the MIR-S slit-less data, which are expected to include fewer point sources than the NIR data due to the long wavelength and low sensitivity in MIR.
The MIR-S channel covers the wavelength range of 5.0--12.8~$\micron$ with two dispersers (SG1: 5.0--8.3~$\micron$; SG2: 7.5--12.8~$\micron$) and a spectral resolution of $R\sim$50.
\citet{Ohyama18} analyzed the MIR-S slit-less data obtained as part of the mission program SPICY (slit-less
SpectroscoPIC surveY of galaxies; PI: T. Wada), which surveyed the north ecliptic pole region.
The present study is the first systematic analysis of the complete MIR-S dataset in the AKARI slit-less data.

\section{Data reduction}

\begin{figure}
 \begin{center}
  \includegraphics[width=0.8\textwidth]{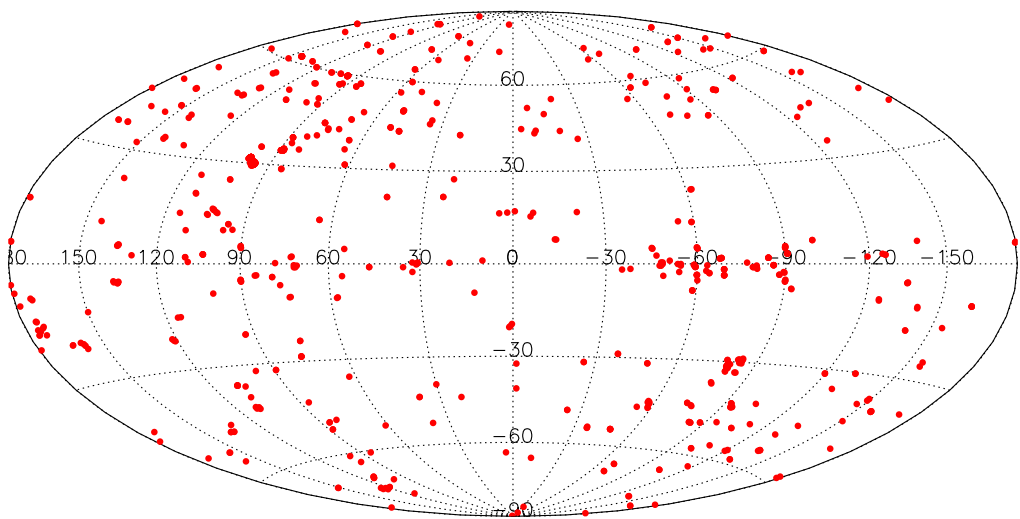} 
 \end{center}
\caption{Positions of all 886 observations obtained by MIR-S slit-less spectroscopy in Galactic coordinates.}\label{obsposition}
\end{figure}

We analyzed all 886 AKARI/IRC MIR-S slit-less data retrievable from the AKARI data archives\footnote{https://darts.isas.jaxa.jp/astro/akari/}, which were carried out with an astronomical observing template of IRC04 for the general spectroscopy.
The observations were made between 2006 Mar. 22 and 2007 Aug. 25 in the cryogenic phase (PV, Phase1, and Phase2).
The MIR-S channel provides spectroscopy by replacing the imaging filters with two transmission-type gratings on the filter wheels (\citealt{Ohyama07}).
Since the FOV of the MIR-S channel is 10$\arcmin \times$10$\arcmin$ (256~pix~$\times$~256~pix, 1~pix = \timeform{2.34"}; \citealt{Onaka07}), the total observed area is 24~deg$^2$, including the overlap.
Figure~\ref{obsposition} shows the positions of the data analyzed in this study.
AKARI/IRC allocates different detector areas to the slit and slit-less spectroscopies, and simultaneously collects data of both modes.
In the cryogenic phase, all spectroscopic channels (NIR, MIR-S, and MIR-L) were operated simultaneously.
Therefore, MIR-S slit-less data were always obtained in the spectroscopic observation, irrespective of the intended observing mode (slit or slit-less) and channel (NIR, MIR-S, or MIR-L).
The simultaneity significantly increases the number of observations in the MIR-S slit-less spectroscopy.

\begin{figure}
 \begin{center}
  \includegraphics[width=0.6\textwidth]{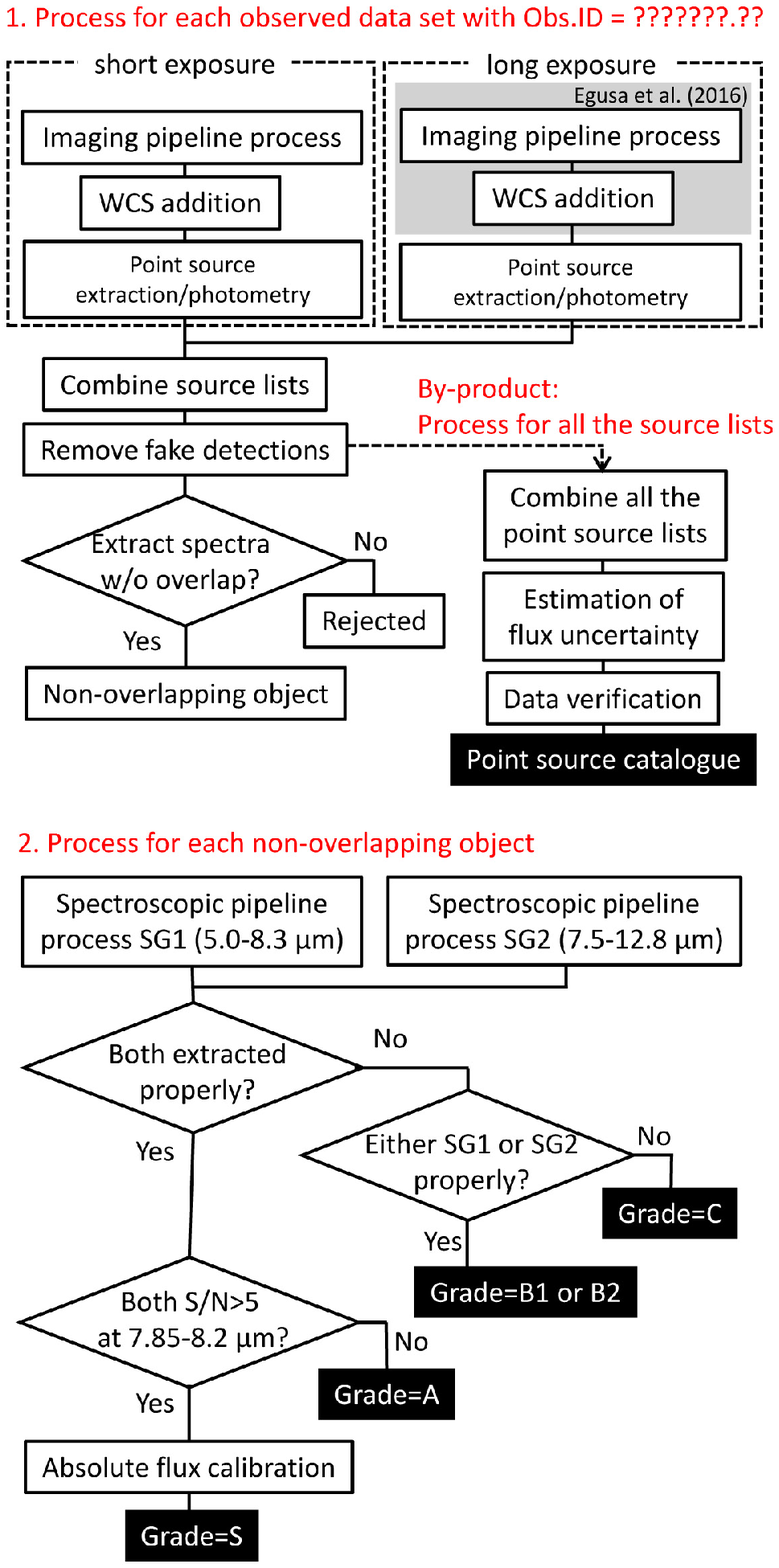} 
 \end{center}
\caption{Flowchart of the data reduction processes.}\label{flowchart}
\end{figure}

Figure~\ref{flowchart} is a flowchart of the overall data reduction.
The data reduction first selects the non-overlapping objects (Step 1), then extracts the slit-less spectra (Step 2).
The reduction begins by analyzing the reference images at wavelength of 9~$\micron$, which were obtained in all spectroscopic observations without a disperser: in the observational sequence, SG1 spectral images, a  reference image with the S9W filter, and SG2 spectral images were obtained in the order, and short- and long-exposure were performed for each image.
We analyzed both short- and long-exposure reference images.
The long-exposure reference images were previously analyzed by \citet{Egusa16}, and the flux-calibrated FITS images with the world coordinate system (WCS) information are available online\footnote{https://www.ir.isas.jaxa.jp/AKARI/Archive/}.
The WCS information was determined by referring to the WISE catalogue (\citealt{Wright10}).
Short-exposure images were analyzed similarly using the IRC imaging toolkit for Phase1\&2 (ver. 20150331).
The point sources in the WCS-added short- and long-exposure reference images were extracted using {\it find.pro} in the IDL Astronomy User's Library, assuming a point-spread function (PSF) size (FWHM) of \timeform{6.0"} (\citealt{Egusa16}).
The detection thresholds of the long- and short-exposure images were 7$\sigma$ and 20$\sigma$, respectively.
The detection threshold was increased for the short-exposure images because only one image was obtained and thus many spurious detections are expected due to the high-energy ionizing particles hit.
From the short-exposure images, we extracted the bright sources that saturate in long-exposure images.
The extracted point sources were processed by PSF photometry and aperture photometry, and the resulting fluxes were used for overlap checking and for accurate absolute flux calibration of the spectra, in the later processes.
The typical fluxes extracted from the long- and short-exposure images were $<$100~mJy and $>$10~mJy, respectively.
In the PSF photometry, the flux was estimated as $F=2\pi \sigma^2 A$, where $\sigma$ is the PSF width ($\sigma$ = \timeform{2.55"} = 2.18~pix, assuming a Gaussian PSF) and $A$ is the peak intensity of the source (output by {\it find.pro}\footnote{We found that the fluxes estimated from short-exposure PSF photometry are systematically overestimated by 10~\% compared with those estimated from the other methods. Since the number of stacked images in short-exposure images is smaller than that in long-exposure images (short: 1 image; long: 3 images), we concluded that the actual PSF size in short-exposure images may be 10\% smaller than that in long-exposure images. Finally, we multiplied the fluxes estimated from short-exposure images by a factor of 0.9 to keep consistency of the fluxes.}).
In the aperture photometry, the source and sky regions were defined as $r<$5~pix and $r=$15--20~pix, respectively, and an aperture correction factor of 0.747 was applied\footnote{https://www.ir.isas.jaxa.jp/AKARI/Observation/support/IRC/ApertureCorrection.html}.
We confirmed that the fluxes obtained from the two methods are consistent within 1\% for spatially isolated objects.
After the photometry, we combined the two point-source lists based on the long- and short-exposure images and made a point source list where duplicated sources with intermediate fluxes (10--100~mJy) are combined.
We removed the fake detections caused by bright objects (\citealt{Arimatsu11}) and hot pixels.
Finally, all sources were checked one by one by visual inspection.
As a result, we obtained 57,968 sources including multiple detections of the same sources, and 42,387 individual sources after combining the multiple detections using a matching radius of \timeform{1.5"}.

\begin{figure}
 \begin{center}
  \includegraphics[width=0.7\textwidth]{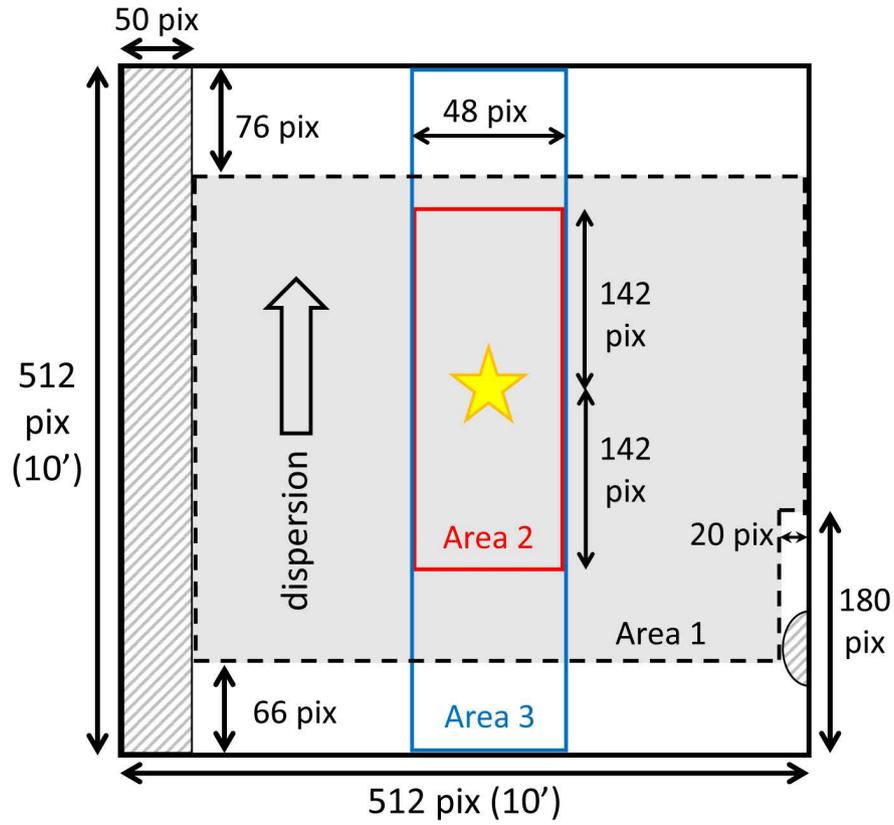} 
 \end{center}
\caption{Schematic of overlap checking on the reference image. Overlaps are checked in Areas 1--3. The light from the objects is dispersed along the vertical direction. The shaded areas are masked in the spectroscopic toolkit.}
\label{check_overlap}
\end{figure}

Figure~\ref{check_overlap} demonstrates the overlap check.
For each point source in turn, the spectral overlap was assessed by following four criteria.
Point sources satisfying all four criteria were regarded as non-overlapping objects, and their spectra were extracted and analyzed in later steps.
\begin{enumerate}
\item The source flux exceeds 1.5~mJy corresponding to the 5$\sigma$ noise equivalent flux of the MIR-S spectroscopy (\citealt{Ohyama07}).

\item The wavelength ranges of SG1 and SG2 spectra are fully covered. To achieve this criterion, the target source must be located in Area 1 of Figure~\ref{check_overlap}, because the light from the source is dispersed in the vertical direction.

\item No other sources with $>$1/20 of the target source flux overlap in the spectrum-extraction area (the 14 pix-wide rectangle on the reference image) and the sky area (12 or 14 pix-wide rectangle on the reference image). This criterion is explained as Area 2 of Figure~\ref{check_overlap}. We first checked the sky areas occupied both sides of the spectrum-extraction area (i.e., the 6~pix-wide area at both sides). If sources other than the target appeared in these areas, we checked cases of one sky area located on one side of the spectrum-extraction area (i.e., the 14~pix-wide area at one side)\footnote{Note that reference images produced from the latest IRC imaging toolkit are applied by sub-pixel sampling. Reference and dispersed images have 512~pix$\times$512~pix and 256~pix$\times$256~pix, respectively, for the same FOV. Parameters used to extract spectra in {\it plot\_spec\_with\_image.pro} are {\it nsum}=7, {\it bg\_offset}=6, {\it bg\_nsum}=3 for the former case,  and {\it nsum}=7, {\it bg\_offset}=8 or -8, {\it bg\_nsum}=7, {\it bg\_oneside}=1 for the latter case.}. Here, objects detected in the MIR were assumed to be dominated by continuum emission. 

\item No bright sources ($>$10~mJy at 9~$\micron$) appear near the target source. This criterion is explained as Area 3 of Figure~\ref{check_overlap}. The zero-th order, second-order, and scattered light from bright sources significantly contaminate nearby spectra in the vertical direction outside of Area 2 (Figure~\ref{check_overlap}). This criterion excludes such contaminations.
\end{enumerate}

In the second step, we extracted the spectra from the non-overlapping sources using IRC spectroscopy toolkit for Phase1\&2 (Release candidate version)\footnote{The updated spectroscopic toolkit used in the present paper will be made available via the AKARI  user support page.}.
Based on the present analysis, we updated the spectral response curves and aperture-correction factors from those of the previous version (ver. 20150331) .
The toolkit updates are summarized in Appendix~1.
We inputted the position of each source on the reference image to the toolkit, and extracted the SG1 and SG2 spectra.
The spectra of sources with $>$1~Jy and $<$1~Jy at 9~$\micron$ were extracted from short-exposure and long-exposure dispersed images, respectively.
Note that the SG1 and SG2 spectra output by the toolkit are sky-subtracted and flux-calibrated in units of mJy.
The major background source in the MIR (apart from the Galactic plane) is the  zodiacal light.
Assuming a 5~MJy/str intensity of the zodiacal light at 9~$\micron$ (\citealt{Kondo16}) and the spectrum-extraction width of 7~pix on the dispersed image, the typical MIR background flux at the ecliptic pole is estimated as $\sim$5~mJy.
If the input position and the actual target position on the dispersed image differed by less than 3~pixels\footnote{We determined the acceptable difference of 3~pixels using a time gap of 1--2 min between the observations of reference images and dispersed images (\citealt{Onaka07}) and the typical drift rate of the telescope of $\sim$2$\arcsec$/min (\citealt{Ohyama07}).} along the cross-dispersion direction, we judged that the spectrum had been extracted properly.
Otherwise, we judged that the variation of the telescope pointing during the observation was not fully corrected.
Such data were discarded because the data were too difficult to correct.
Note that we did not consider the positional difference along the dispersion direction.
If both the SG1 and SG2 spectra were extracted properly and detected with S/N$>$5 in the overlapping wavelength (7.85--8.2~$\micron$), we calibrated their absolute fluxes (flux recalibration) to maintain consistency between the photometric S9W flux estimated from the aperture photometry and the spectroscopic S9W flux estimated from the SG1/SG2 spectra and the relative spectral response curve of the S9W band.
After the absolute flux calibration, the SG1 and SG2 spectra were scaled and combined into a spectrum.
The wavelength range of the combined spectrum was limited to 5.5--12.5~$\micron$ to avoid the large uncertainty caused by the low optical throughput near the edges of the wavelength coverage (the wavelength coverage of the original SG1 and SG2 spectra were 5.0--8.3~$\micron$ and 7.5--12.8~$\micron$, respectively).
As a result, we extracted 862 spectra of 604 objects including multiple detections.
Overall, 1.5\% of the MIR spectra were extracted to the point source lists (i.e., 862 out of 57,968 sources).

Each extracted spectrum was graded by a quality flag: S (490 spectra), A (270 spectra), B1 (71 spectra), B2 (25 spectra), or C (7 spectra) depending on its extraction condition.
The flux-recalibrated spectra were S-graded, indicating sufficient quality for scientific analysis.
If either the SG1 or SG2 spectrum was undetected with S/N$>$5 in the 7.85--8.2~$\micron$ range, the spectrum was assigned an A grade.
We recommend that users use A-graded spectra for scientific analysis after careful consideration of the absolute flux levels. 
If either the SG1 or SG2 spectrum was properly extracted, the spectrum was assigned a B1 or B2 grade, respectively.
For B1 (B2)-graded spectra, the SG1 (SG2) spectrum can be used for the scientific purpose with a caveat on the absolute flux level as A-graded spectra, while the SG2 (SG1) spectrum is not recommended to use because the spectrum extraction has probably failed due to unstable telescope pointing.
If neither SG1 nor SG2 spectra are properly extracted, the spectrum was assigned a C grade (such spectra are not recommended for scientific use).
Note that only S-graded spectra have high absolute flux accuracy ($\sim$6\%).
Only toolkit-output SG1 and SG2 spectra are available for A-, B1-, and B2-graded spectra, which have relatively low absolute flux accuracy (8--14~\%).
The absolute flux calibration of spectra is discussed in subsection~4.1.

\begin{figure}
 \begin{center}
  \includegraphics[width=0.7\textwidth]{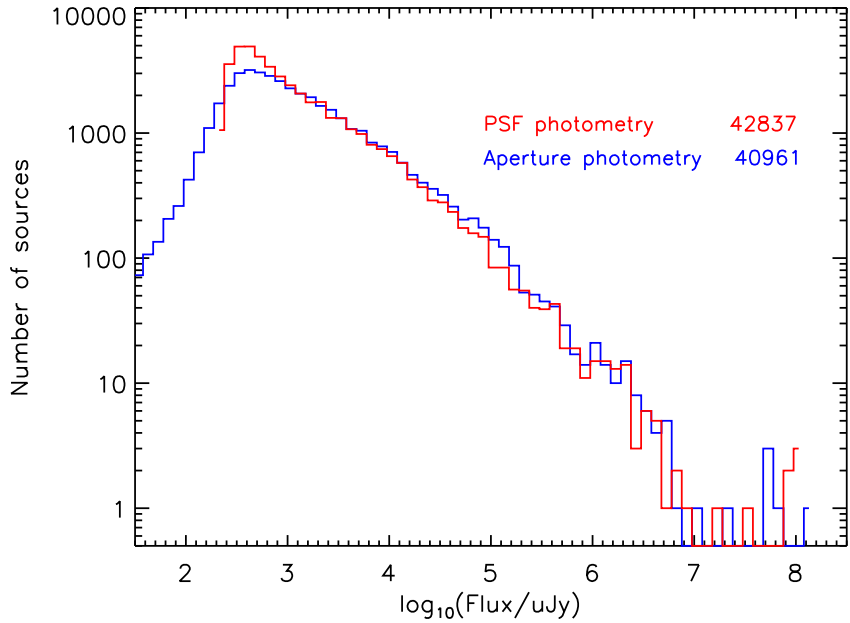} 
 \end{center}
\caption{Histogram of the 9~$\micron$ fluxes in the point-source catalogue. The fluxes were estimated by PSF photometry (red) and aperture photometry (blue). The numbers at the upper right are the total numbers of sources measured in the two methods.}\label{flux_histogram}
\end{figure}

We also constructed a by-product catalogue that combines all the point-source lists made in the first step.
After combining the multiple detections in the 57,968 sources using a \timeform{1.5"} matching radius, we obtained a point-source catalogue of 42,387 sources.
If a source was detected twice or more, its flux was averaged to obtain the catalogue entry.
Figure~\ref{flux_histogram} shows histograms of the 9~$\micron$ fluxes (from PSF and aperture photometry data) in the point-source catalogue.
Sources brighter than $\sim$0.3~mJy are included in the catalogue.
The 5$\sigma$ point-source sensitivities of the WISE W3 band are $\sim$0.9~mJy and $\sim$0.5~mJy near the Galactic plane and pole, respectively (\citealt{Wright10}).
Therefore, the present point-source catalogue is 2--3 times deeper than the WISE catalogue in the MIR.
The estimated flux uncertainty, accuracy of the absolute flux calibration, and positional accuracy of the point-source catalogue are presented in Appendix~2.

The spectral and point-source catalogues are available online\footnote{https://www.ir.isas.jaxa.jp/AKARI/Archive/}.
The catalogues are formatted as shown in Appendix~3.
The spectral catalogue includes the toolkit-output spectra (all sources) and flux-recalibrated spectra (Grade=S only) in ASCII format, and the supplemental data (quick-look images made by the spectroscopic toolkit).
The point-source catalogue lists the 9~$\micron$ flux, number of detections, number of other sources near the target, and the corresponding sources in the 2MASS and WISE catalogues.
All reference images from which we extracted the point sources are also available.

\section{Results}

\begin{figure}
 \begin{center}
  \includegraphics[width=0.8\textwidth]{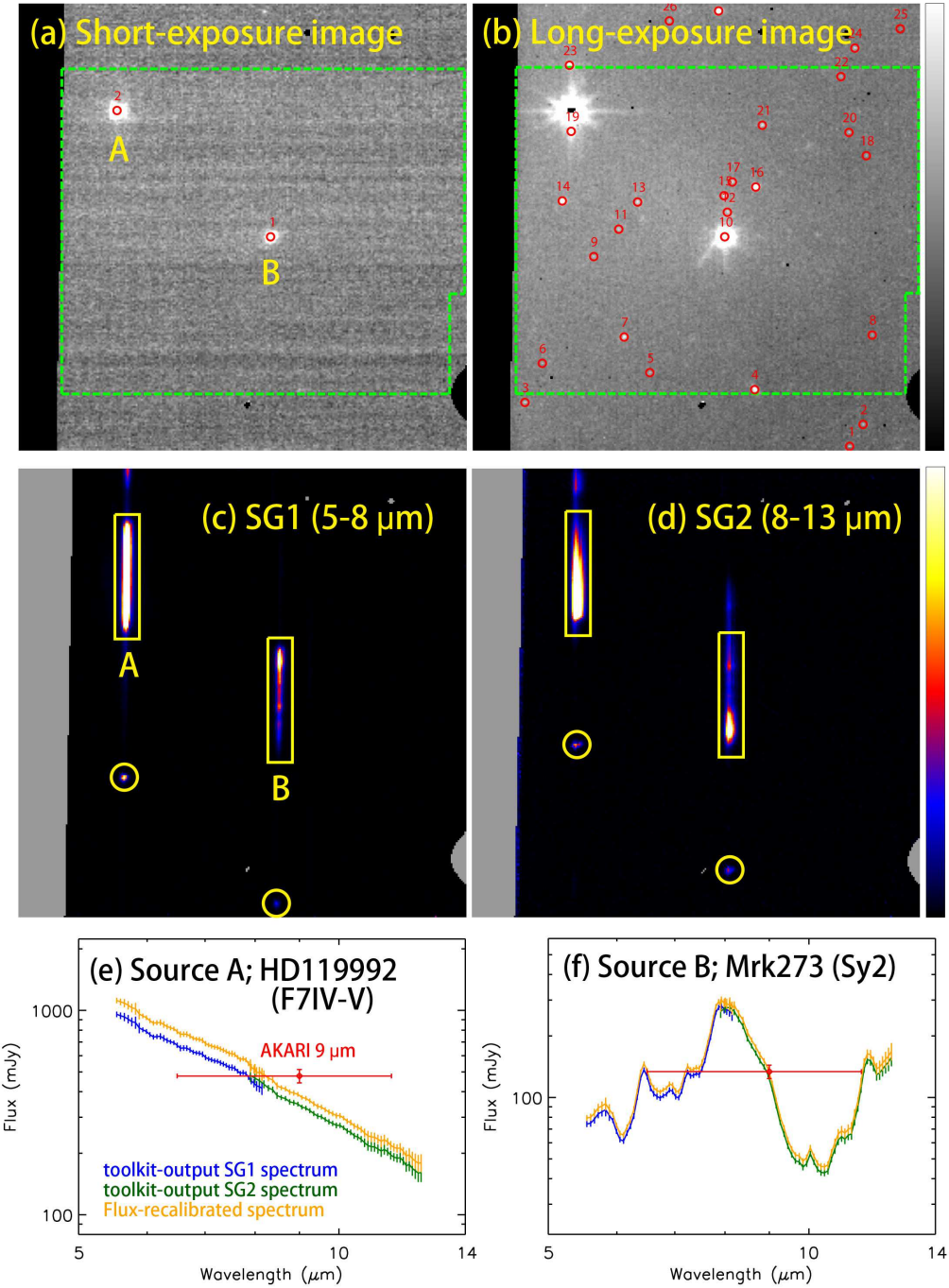} 
 \end{center}
\caption{Example of data analysis in an observed region (ObsID = 1100298.1). (a, b) Point source extraction from short- and long-exposure reference images. Red circles enclose positions of the extracted point sources. The green box corresponds to Area 1 in Fig.~\ref{check_overlap}. (b, c) Spectral images with two dispersers, SG1 and SG2. The spectra were extracted from the regions enclosed in the boxes. The circles enclose the zero-th order light of sources A and B. (e,f) Spectra extracted from sources A and B. The blue and green curves plot the SG1 and SG2 spectra, respectively, output from the spectroscopic toolkit, while the orange curve indicates the final spectrum after the absolute flux calibration using the AKARI 9~$\micron$ flux (red data point).}\label{data_example}
\end{figure}

Figure~\ref{data_example} demonstrates the data analysis in an observed region (ObsID = 1100298.1).
In this region, 2 and 27 point sources were identified in the short- and long-exposure reference images (panels (a) and (b)), respectively.
Sources A and B were confirmed as non-overlapping objects by the overlap check.
Owing to saturation, the long-exposure reference image of the brighter source A is unavailable for analysis.
This demonstrates that combining the short- and long-exposure reference images covers a sufficiently wide flux range for the spectra extraction.
From sources A and B within the yellow boxes in panels (c) and (d), we extracted the SG1 and SG2 spectra and display them in panels (e) and (f).
To obtain the flux-recalibrated spectra, we multiplied the toolkit-output spectra by a factor of $\sim$1.1.
Whereas the main target of the observation was source B (Mrk~273), we obtained not only the spectrum of B, but also a spectrum from source A (HD~119992).
Such serendipitous detection is one strength of slit-less spectroscopy.

\begin{figure}
 \begin{center}
  \includegraphics[width=\textwidth]{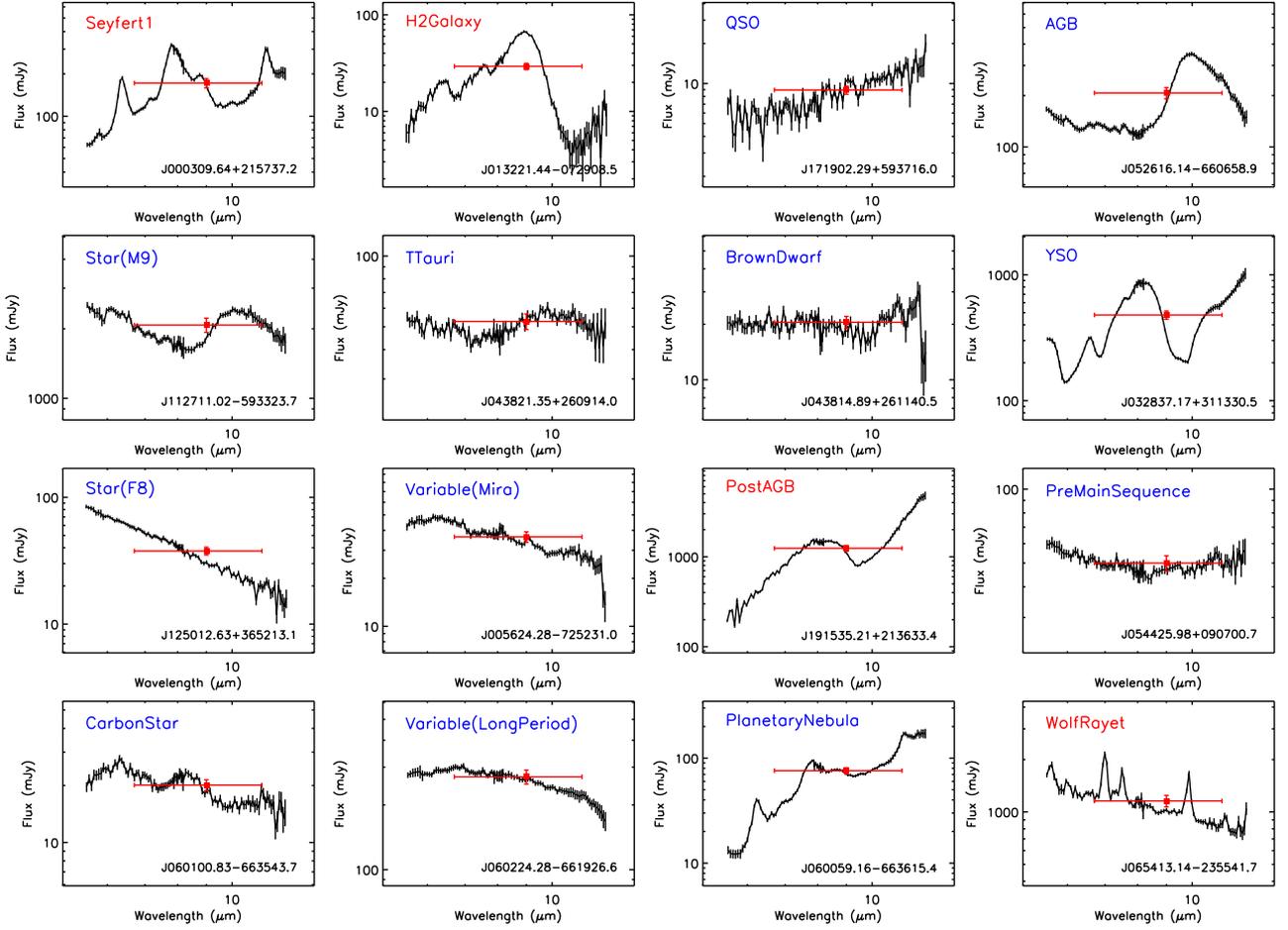} 
 \end{center}
\caption{Examples of S-graded spectra. Top left and bottom right labels in each panels are the object classifications by SIMBAD and the object IDs in the 9~$\micron$ point-source catalogue, respectively. The red font refers to the objects targeted by the observer, while the blue font indicates the serendipitous objects detected in our analysis.}\label{spec_list}
\end{figure}

Figure~\ref{spec_list} shows examples of the S-grade spectra extracted in our analysis.
As evidenced in the figure, a variety of spectra are stored in the spectral catalogue.
We checked classification of all 604 objects in SIMBAD with a matching radius of \timeform{1.5"} (see Table~\ref{classification}).
The catalogue includes many types of objects, from stars to galaxies, and 52\% of the objects (315 out of 604) have no SIMBAD identifications.
We also performed cross-matching between the present spectroscopic catalogue, the 2MASS point source catalogue (\citealt{Skrutskie06}), and the WISE all-sky source catalogue (\citealt{Wright10}), and found that 96\% of all objects (578 out of 604) were matched to a counterpart within \timeform{1.5"} either in the 2MASS or WISE catalogue.
That is, a large fraction of the non-identified objects in the present catalogue have photometric data, but their natures have not been identified.
By matching the coordinates of the extracted objects to those originally requested by the observer (with a matching radius of \timeform{1.5"}), we also checked whether each object was the intended object of the observer.
We found 26 originally intended spectra and 836 serendipitously detected spectra (labeled red and blue in Figure~\ref{spec_list}, respectively).
This analysis confirmed the predominance of serendipitous detections in the present spectral catalogue.

\begin{figure}
 \begin{center}
  \includegraphics[width=0.65\textwidth]{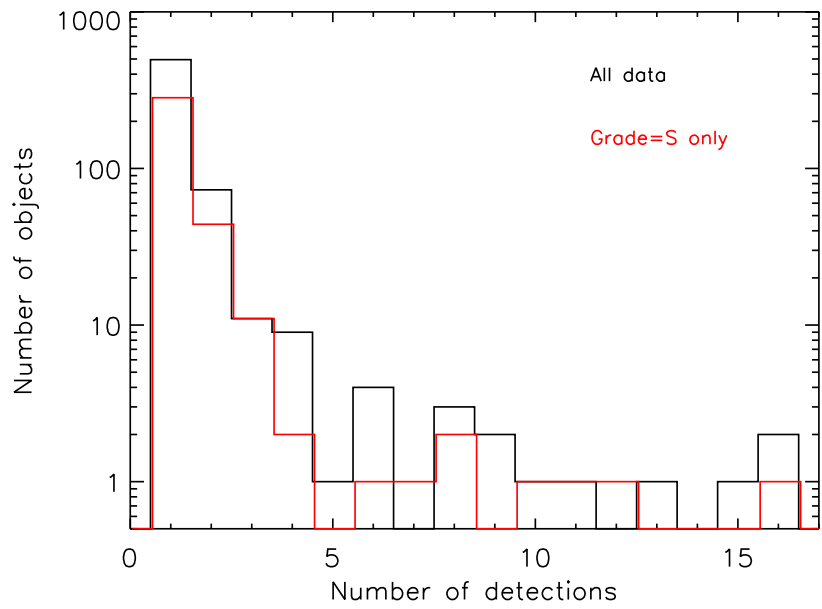} 
 \end{center}
\caption{Histograms of number of detections in the spectral catalogue. The red histogram is the result of the S-graded spectra only.}\label{histogram_spec}
\end{figure}

Figure~\ref{histogram_spec} shows the histograms of the numbers of detections in the spectral catalogues of all spectra and S-graded spectra.
As can be seen in the figure, the number of spectra for each object ranges from 1 to 16, and 80~\% of the objects have only one available spectrum.
Spectra of the same object are not averaged in the spectral catalogue and all extracted spectra are provided as separate data.
Therefore, users are recommended to check the reliability of the spectra by comparing the multiple spectra when available.
Figure~\ref{stability} shows the spectra of ten S-grade objects detected 4--16 times.
The upper parts of each panel superimpose the spectra of the same object, while the lower parts present the ratio of the standard deviation and average flux, which corresponds to the signal-to-noise ratio of the spectra.
The spectra are matched within the noise level, confirming the reliability of the extraction process.
We confirmed that differences among the spectra of the same object can be attributed to the fluctuating sky signals (Appendix~4).
Roughly speaking, a spectrum brighter than 10~mJy matches other spectra of the same object within 10~\%.
The relatively low signal-to-noise ratio at wavelengths above 10~$\micron$ is expected from the low sensitivity in the spectral range (\citealt{Ohyama07}).

\section{Discussion}

\subsection{Accuracy of the absolute flux calibration}

This section checks the absolute flux accuracy of grade-S spectra and the lower-grade spectra.
The absolute flux of the grade-S spectra was guaranteed in comparisons between the photometric and spectroscopic S9W fluxes.
Therefore, the absolute flux accuracy of these spectra is closely linked to the photometry accuracy.
\citet{Tanabe08} estimated the absolute flux calibration accuracy of AKARI/IRC as 6\% in the S9W band.
To evaluate the flux calibration, we compared the 9~$\micron$ fluxes measured from the reference image with those in the AKARI/IRC point-source catalogue (\citealt{Ishihara10}).
Since the same S9W filter was used in the reference image and the AKARI all-sky survey but the pointed observation and the AKARI all-sky survey were independently calibrated, these fluxes are directly comparable.
We performed cross-matching with a matching radius of \timeform{1.5"}.
Figure~\ref{flux_calibration_GradeS}(a) compares the fluxes estimated from the reference image and those retrieved from the AKARI all-sky point-source catalogue.
The fluxes are consistent within 3~\%, confirming the consistency of the absolute flux calibrations of the S9W band and the MIR-S slit-less spectra.

It should be noted that owing to the low sensitivity of the AKARI all-sky survey, the above comparison is valid only for bright sources ($>$50~mJy or $<$7.3~mag at 9~$\micron$).
Therefore, in the same manner, we also compared the fluxes measured from reference images with those in the WISE all-sky source catalogue (\citealt{Wright10}).
Since the central wavelength of the WISE W3 band is 12~$\micron$, we compared the fluxes in the magnitude scale only for stars identified by cross-matching with SIMBAD (Table~\ref{classification}).
As shown in Figure~\ref{flux_calibration_GradeS}(b), the magnitudes at 9~$\micron$ and 12~$\micron$ were consistent within 6\%.
Moreover, the comparison held over a wide flux range (2--11~mag or 2~mJy--9~Jy).
The consistency validates the absolute flux calibration of the slit-less spectra.

The absolute fluxes of the A-, B1-, and B2-graded spectra are not recalibrated, owing to the low S/N ratio or incomplete wavelength coverage.
For these spectra, only toolkit-output SG1 and SG2 spectra are provided.
To enable use of A-, B1-, and B2-graded spectra, we indirectly estimated their absolute flux accuracies by using the absolute flux calibration factors applied to the toolkit-output SG1 and SG2 spectra, thereby forming grade-S spectra.
Figure~\ref{histogram_scaling} shows histograms of the absolute flux calibration factors.
As evidenced in the figure, the scaling factors typically exceed 1.0; average values of 1.14 and 1.08 with 1$\sigma$ standard deviations of 0.17 and 0.17 for the toolkit-output SG1 and SG2 spectra, respectively.
Hence, when using the A-, B1-, and B2-graded spectra without considering their absolute flux levels, their relatively large uncertainties should be also recognized.

\subsection{Accuracy of wavelength}

The wavelength accuracy was verified by measuring the wavelengths of known emission lines.
Preferably, the wavelength accuracy should be systematically testified on many objects.
However, many of the objects in the spectral catalogue are various types of stars (Table~\ref{classification}) with no emission lines in the MIR.
For this analysis, we selected three objects (HD~50896, Mrk~3, and NGC~6240) with clear emission lines in both the SG1 and SG2 ranges.
After fitting Gaussian functions to the emission lines, we measured the differences between the observed and expected wavelengths.
The continuum emission was reproduced by a power-law function or a first-order polynomial.
Panels (a)--(c) of Figures~\ref{wavelength_check} show the fitting results and the identified emission lines used in the analysis, and panel (d) summarizes the differences between the observed and expected wavelengths.
The observed and expected wavelength differed by (at most) 0.06~$\micron$ in the SG1 range and 0.15~$\micron$ in the SG2 range.
Since the detector dispersion is 0.057 and 0.097 $\micron$/pix in SG1 and SG2, respectively (\citealt{Ohyama07}), the observed wavelength accuracy is $\sim$1.5~pix at worst.
\citet{Ohyama07} estimated the typical overall wavelength accuracy as 1~pix.
Hence the wavelength accuracy in the slit-less spectroscopy is consistent with earlier estimates of \citet{Ohyama07}.

\subsection{Comparison with Spitzer/IRS spectra}

Figure~\ref{spitzer_hikaku} compares the AKARI/IRC slit-less spectra and Spitzer/IRS slit spectra of commonly observed stars and galaxies.
The Spitzer/IRS spectra were retrieved from the CASSIS archives (\citealt{Lebouteiller11}).
Since the Spitzer/IRS spectra were obtained with a \timeform{3.6"}-wide SL slit, they may miss a part of the total flux.
For this reason, we scaled the Spitzer/IRS spectra to match the AKARI/IRC slit-less spectra at 8.0~$\micron$, and examined their spectral-shape differences only.
As evidenced in Figure~\ref{spitzer_hikaku}, the AKARI/IRC slit-less spectra and Spitzer/IRS slit spectra are generally consistent within the errors.
Spitzer/IRS slit spectra of galaxies tend to show a systematically deeper absorption around $\sim$10~$\micron$ than the corresponding AKARI/IRC slit-less spectra, which is attributed to silicate.
Differences may be caused by the different aperture sizes of the spectral extraction; AKARI/IRC sums up all signals from the object using the equal weight, while Spitzer/IRS sums up signals with the weight of the PSF profile (\citealt{Lebouteiller11}).
Consequently, Spitzer/IRS spectra are more sensitive to the central part of the object.
Since the commonly observed galaxies are (ultra) luminous infrared galaxies and active galactic nuclei, the dense ISM may reside in the galactic center.
Therefore, the Spitzer/IRS spectra, focusing on the galactic center, may show systematically deeper absorption than the AKARI/IRC spectra.
We conclude that the AKARI/IRC and Spitzer/IRS provide consistent spectra.

\section{Summary}

We have analyzed all 886 AKARI/IRC MIR-S slit-less spectroscopic data obtained in the cryogenic phase.
We constructed a point-source list in each FOV and checked their spectral overlaps using their relative positions and fluxes.
We have obtained 862 non-overlapping MIR spectra in the 5--13~$\micron$ range from 604 individual objects and photometric data of 42,387 objects at 9~$\micron$ as a by-product.
The slit-less spectra include a variety of objects, from stars to galaxies.
We have verified the quality of the spectra and estimated the absolute flux accuracies as 6~\% in the flux-recalibrated spectra and 8--14~\% in the toolkit-output SG1 and SG2 spectra.
The wavelength accuracy is $\sim$0.1~$\micron$.
Moreover, the AKARI/IRC slit-less spectra and the Spitzer/IRS slit spectra are satisfactorily consistent.
The spectroscopic and point source catalogues are available online.


\begin{ack}
We express many thanks to the anonymous referee for useful comments.
We thank all the members of the AKARI project.
This research is based on observations with AKARI, a JAXA project with the participation of ESA.
This research has made use of the SIMBAD database, operated at CDS, Strasbourg, France.
This work is also based on observations made with the Spitzer Space Telescope, which is operated by the Jet Propulsion Laboratory, California Institute of Technology under a contract with NASA.
This publication makes use of data products from the Wide-field Infrared Survey Explorer, which is a joint project of the University of California, Los Angeles, and the Jet Propulsion Laboratory/California Institute of Technology, funded by the National Aeronautics and Space Administration.
MY is supported by JSPS KAKENHI Grant Number JP17K14261.
\end{ack}

\begin{figure}
  \includegraphics[width=0.24\textwidth]{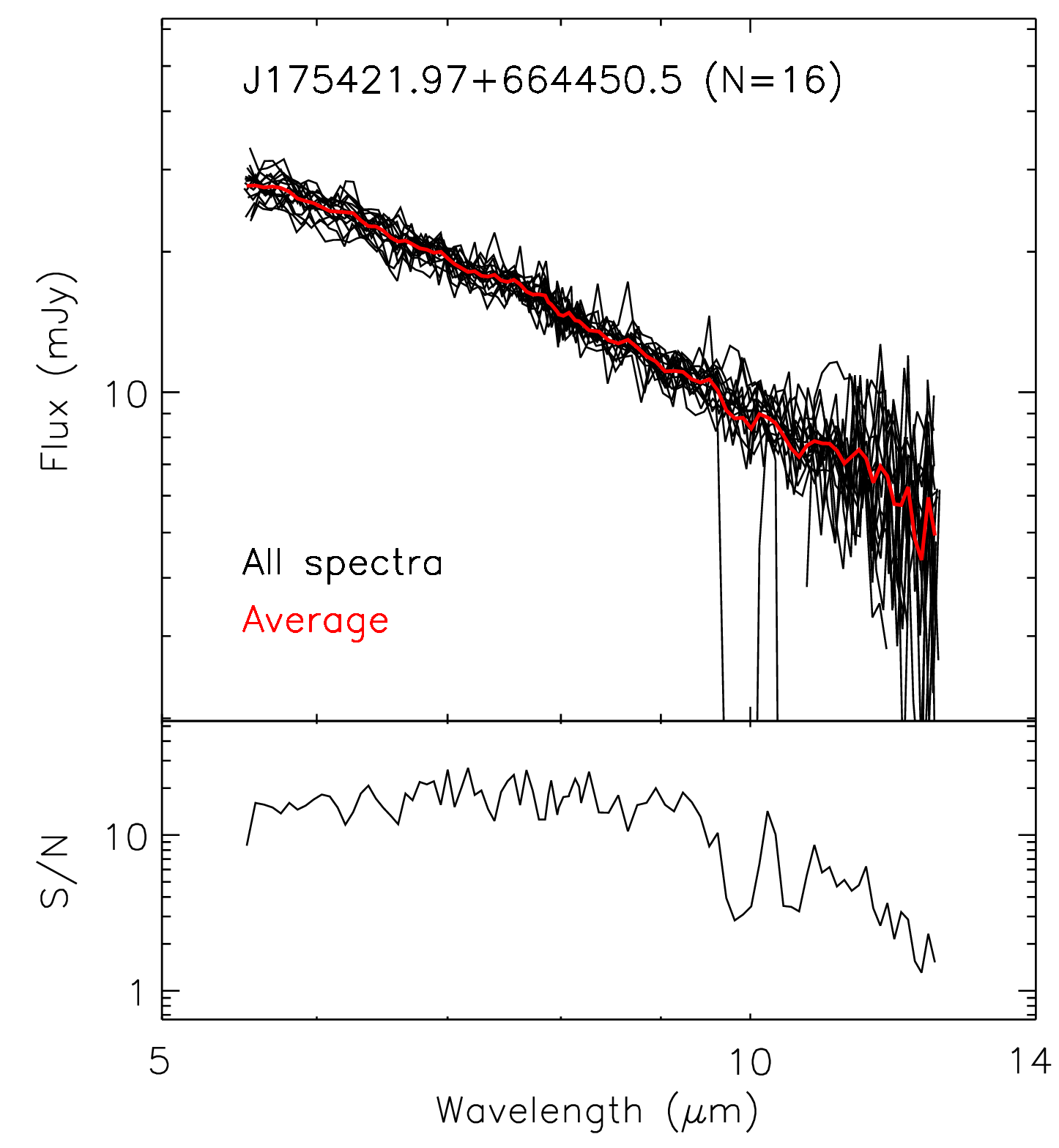} 
  \includegraphics[width=0.24\textwidth]{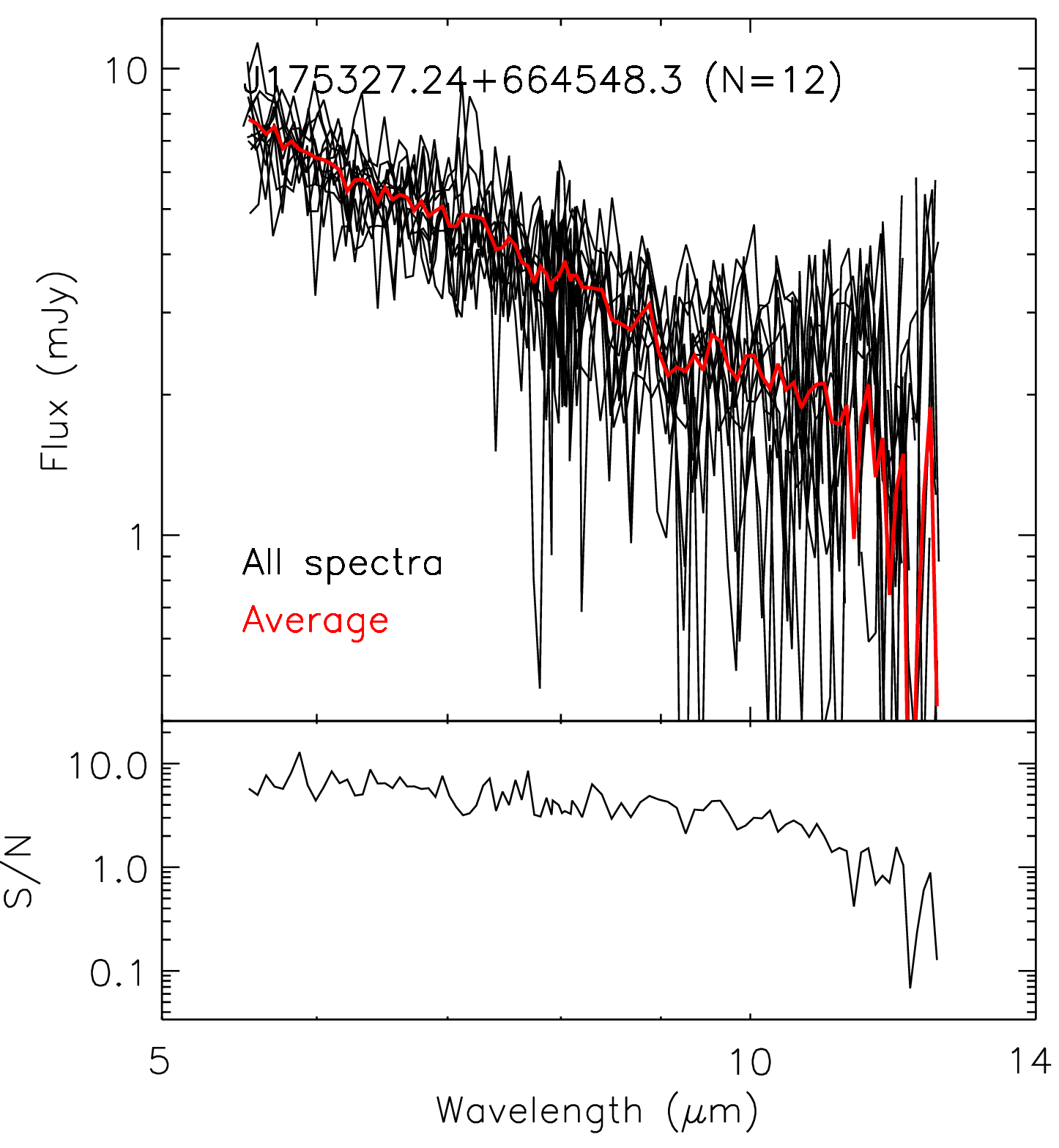} 
  \includegraphics[width=0.24\textwidth]{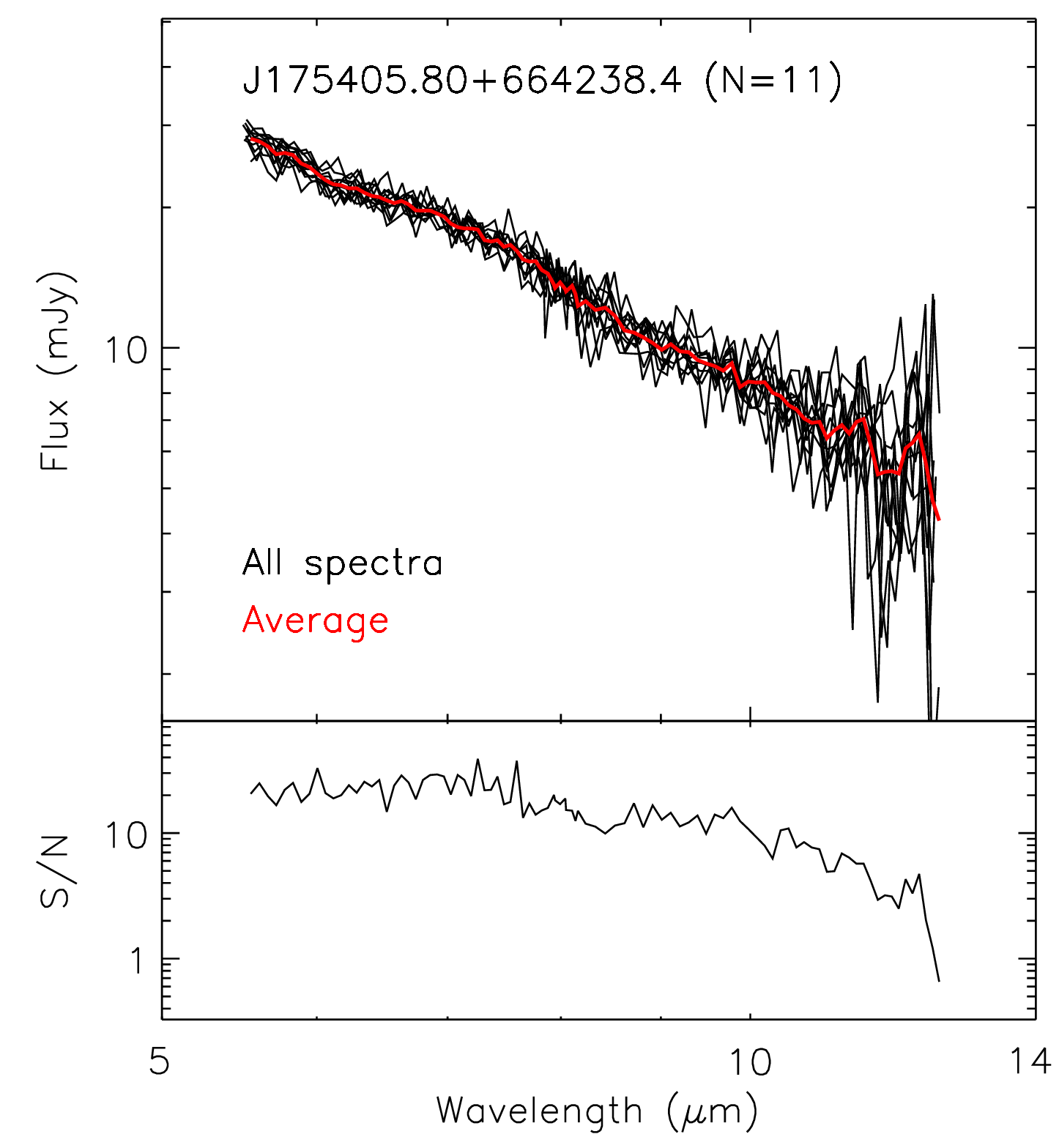} 
  \includegraphics[width=0.24\textwidth]{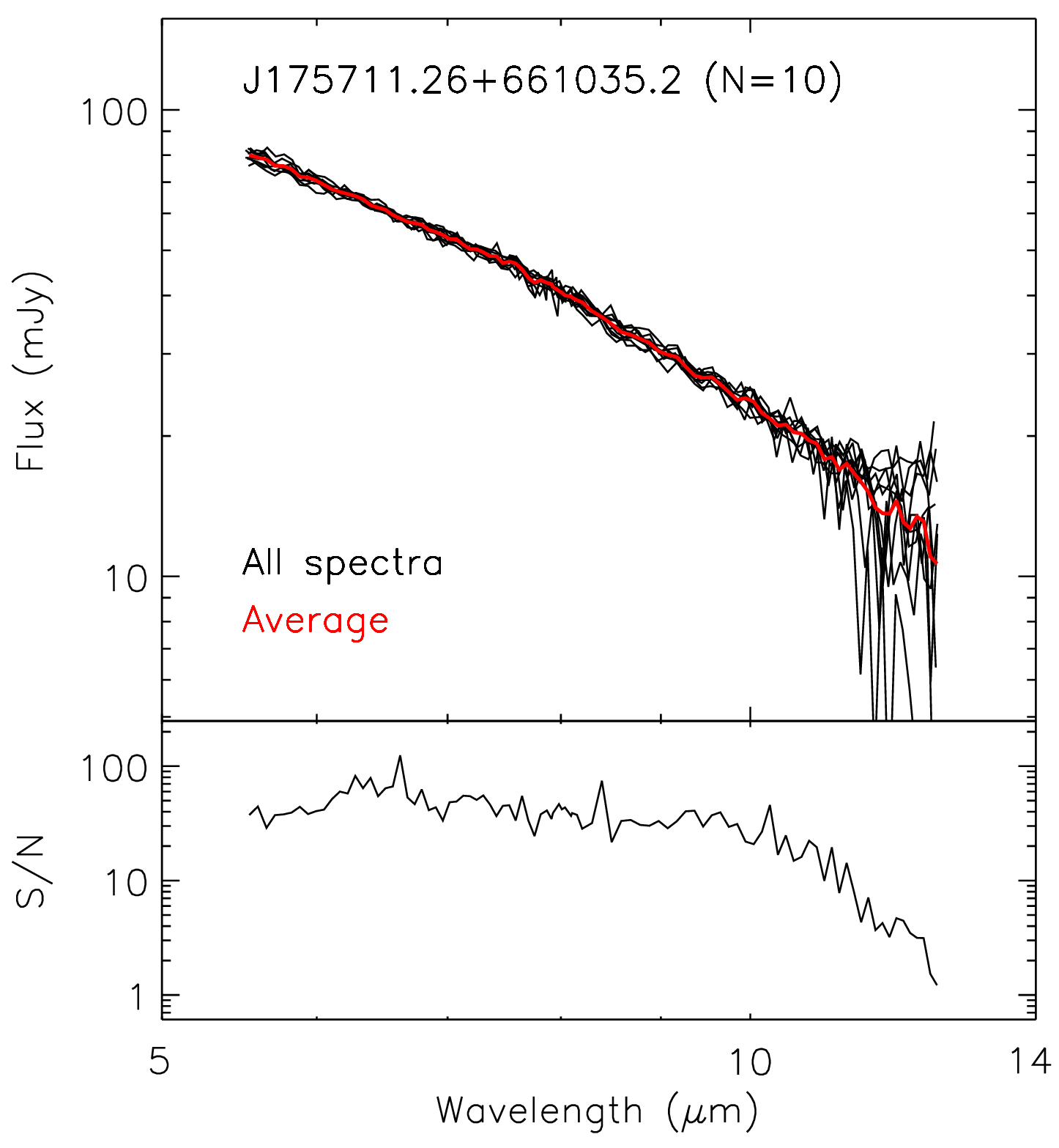} \\
  \includegraphics[width=0.24\textwidth]{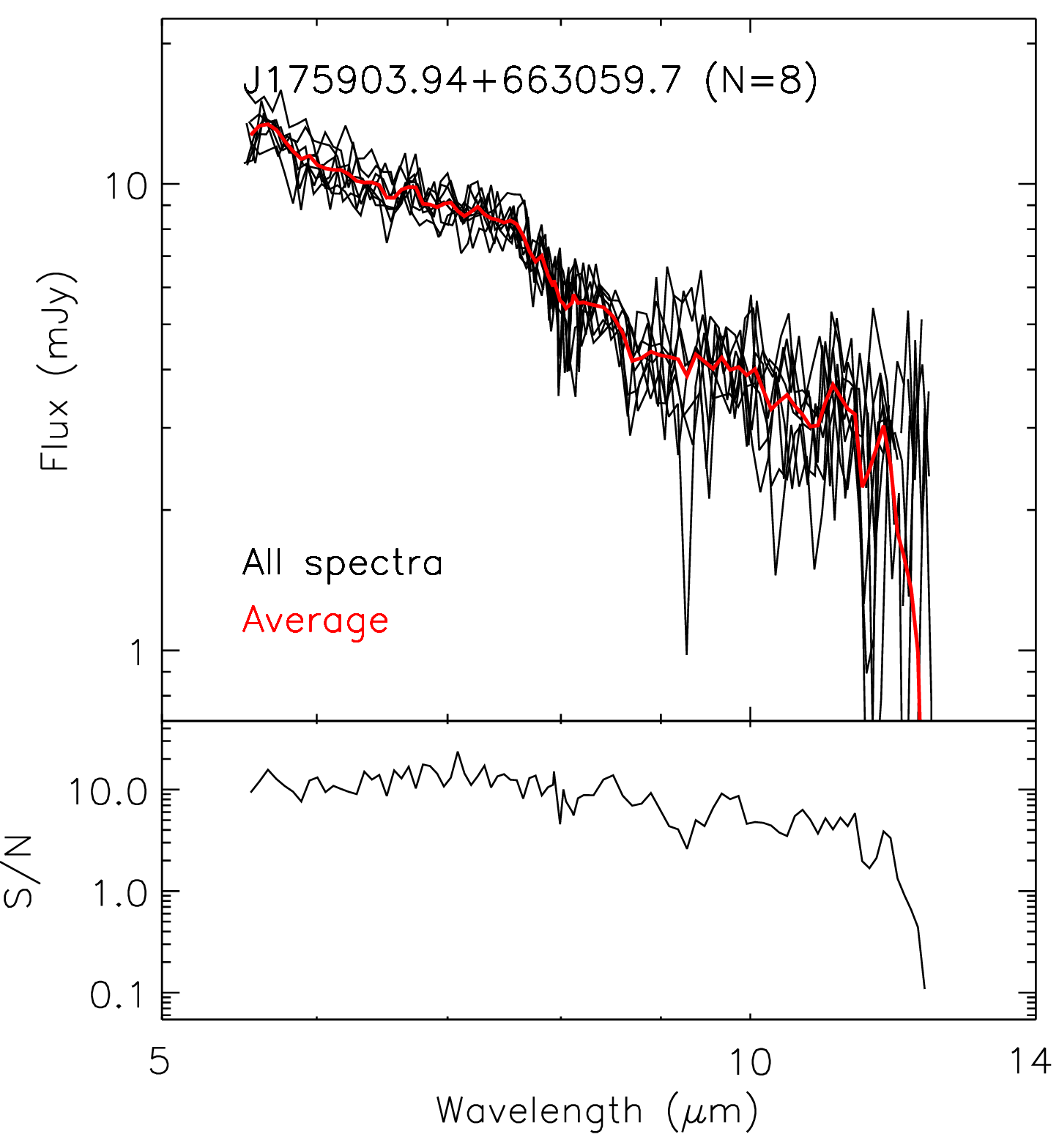} 
  \includegraphics[width=0.24\textwidth]{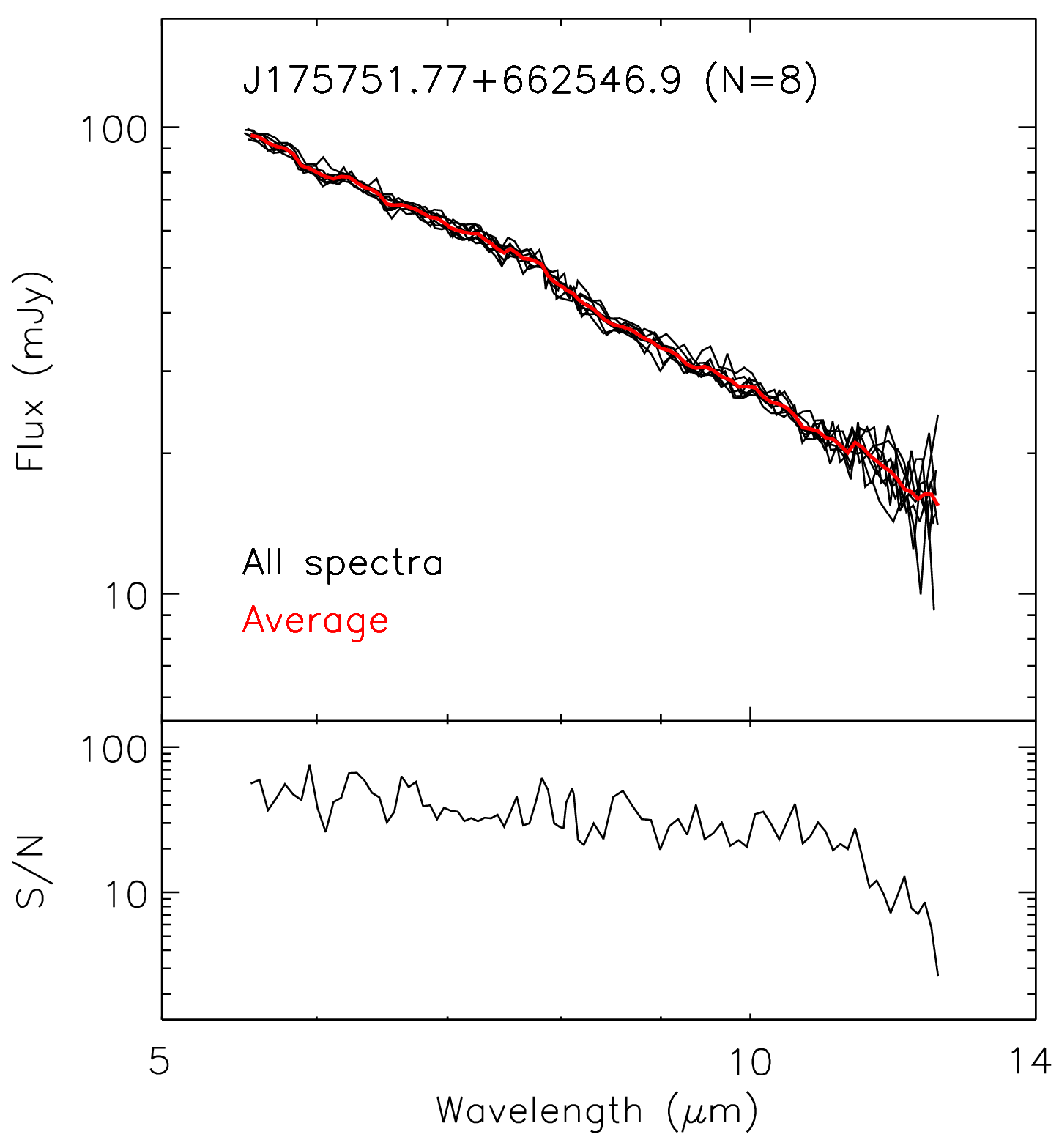} 
  \includegraphics[width=0.24\textwidth]{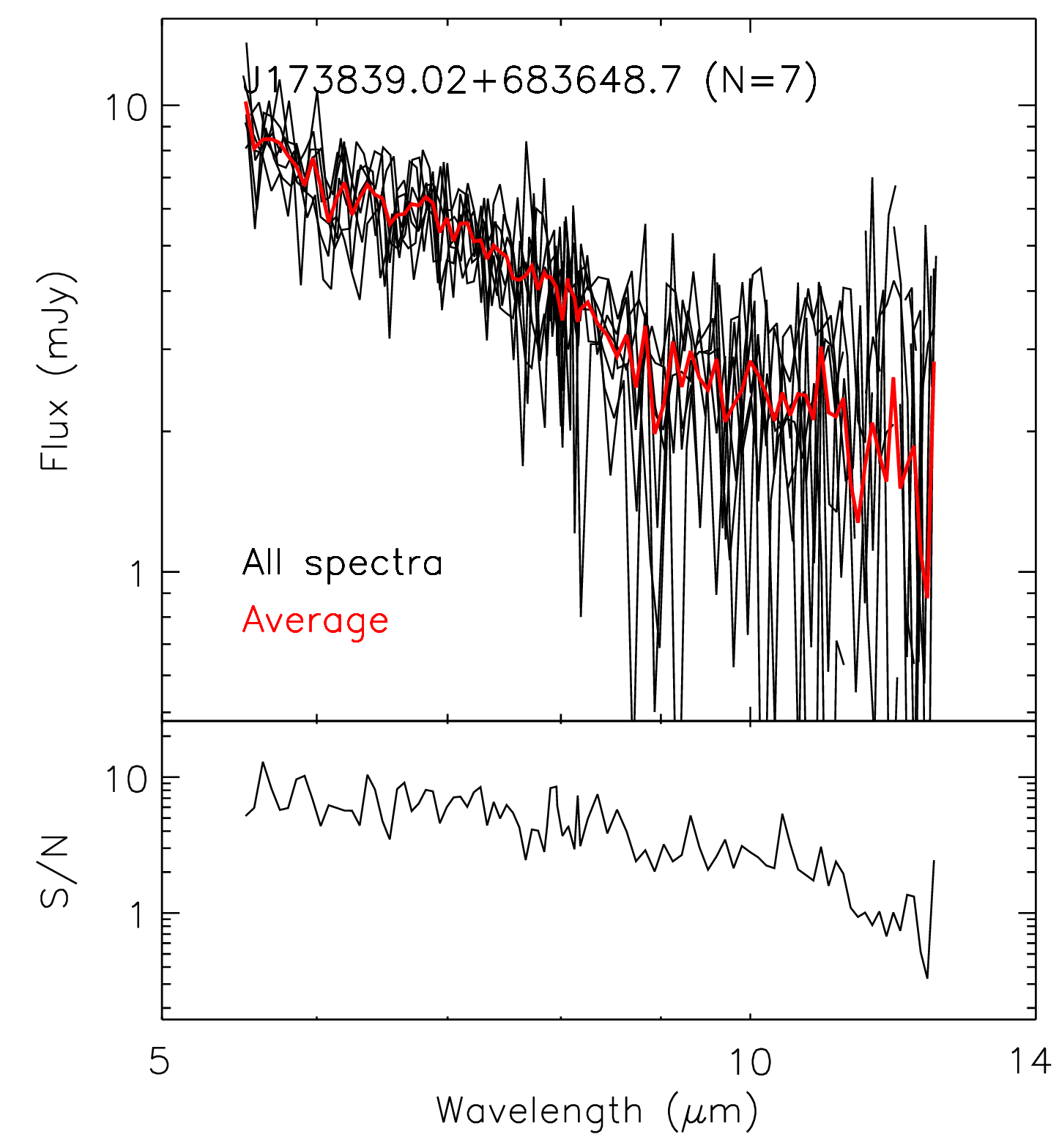} 
  \includegraphics[width=0.24\textwidth]{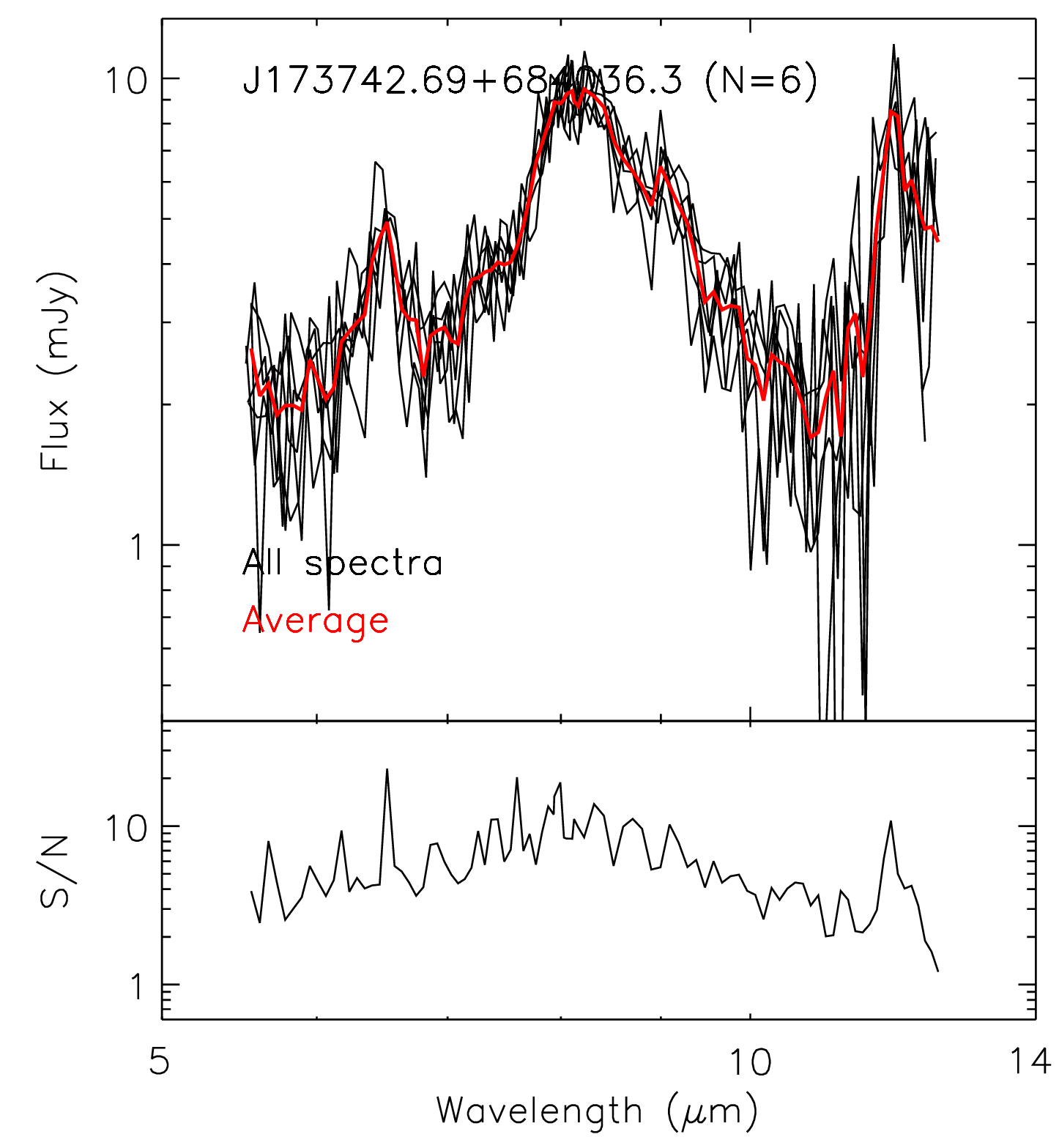} \\
  \includegraphics[width=0.24\textwidth]{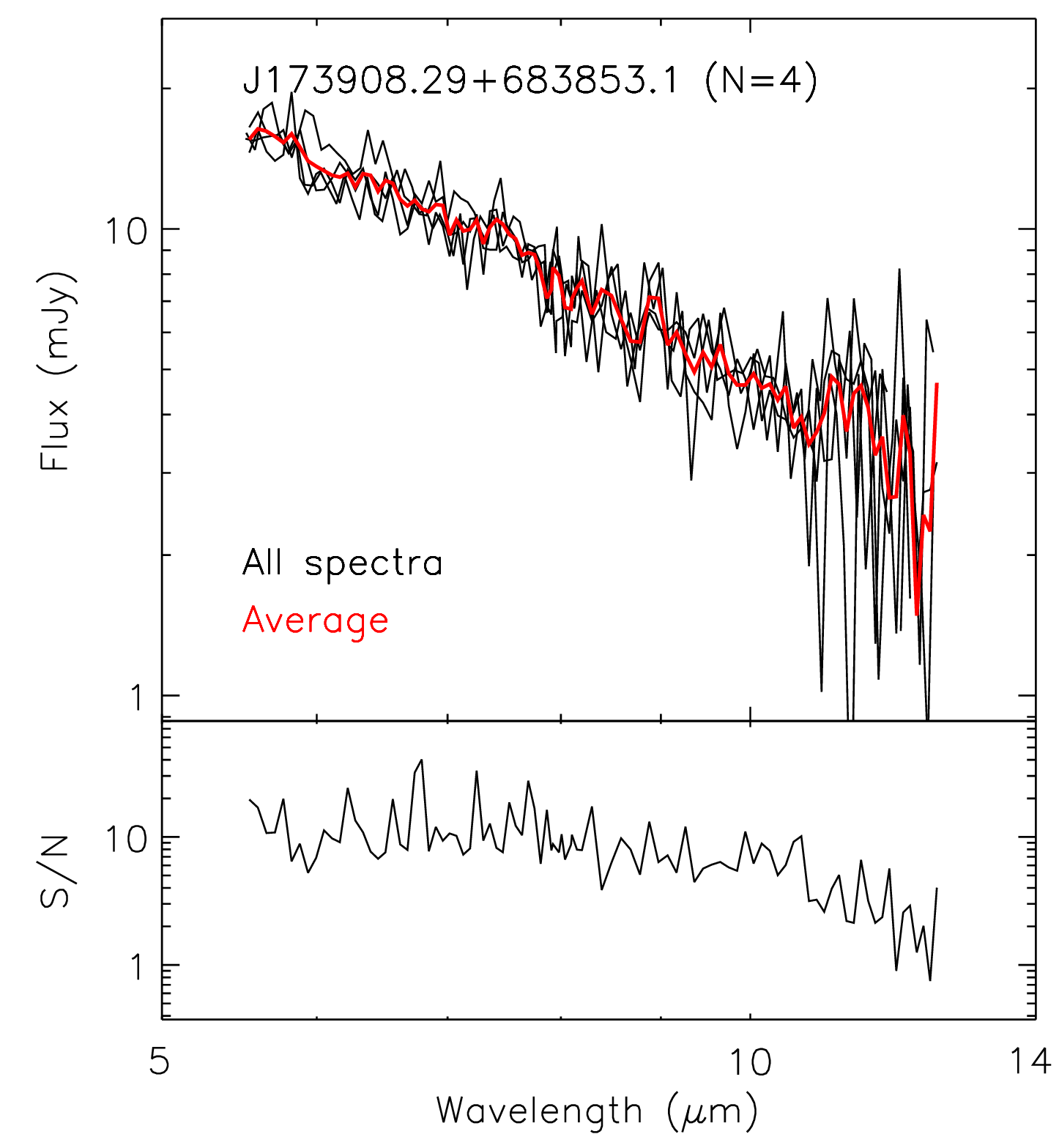} 
  \includegraphics[width=0.24\textwidth]{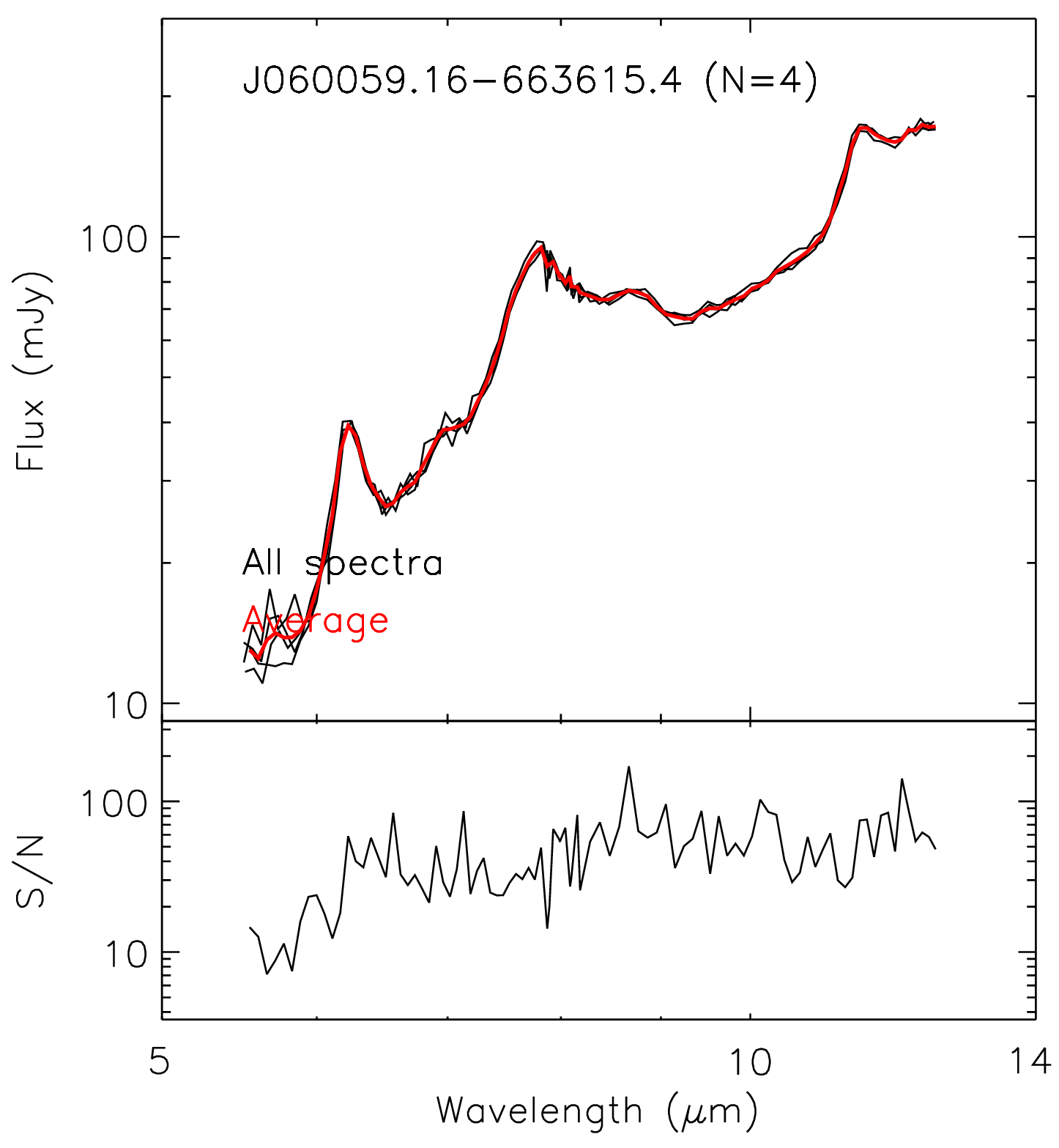} 
\caption{(Upper parts) Superimposed S-graded spectra detected four times or more. Red curves are the averaged spectra. Object ID and the number of grade-S detections are shown at the top of each panel. (Lower parts) Ratio of the standard deviation to the average fluxes in each wavelength bin (S/N spectra).}\label{stability}
\end{figure}

\begin{figure}
 \begin{center}
  \includegraphics[width=0.48\textwidth]{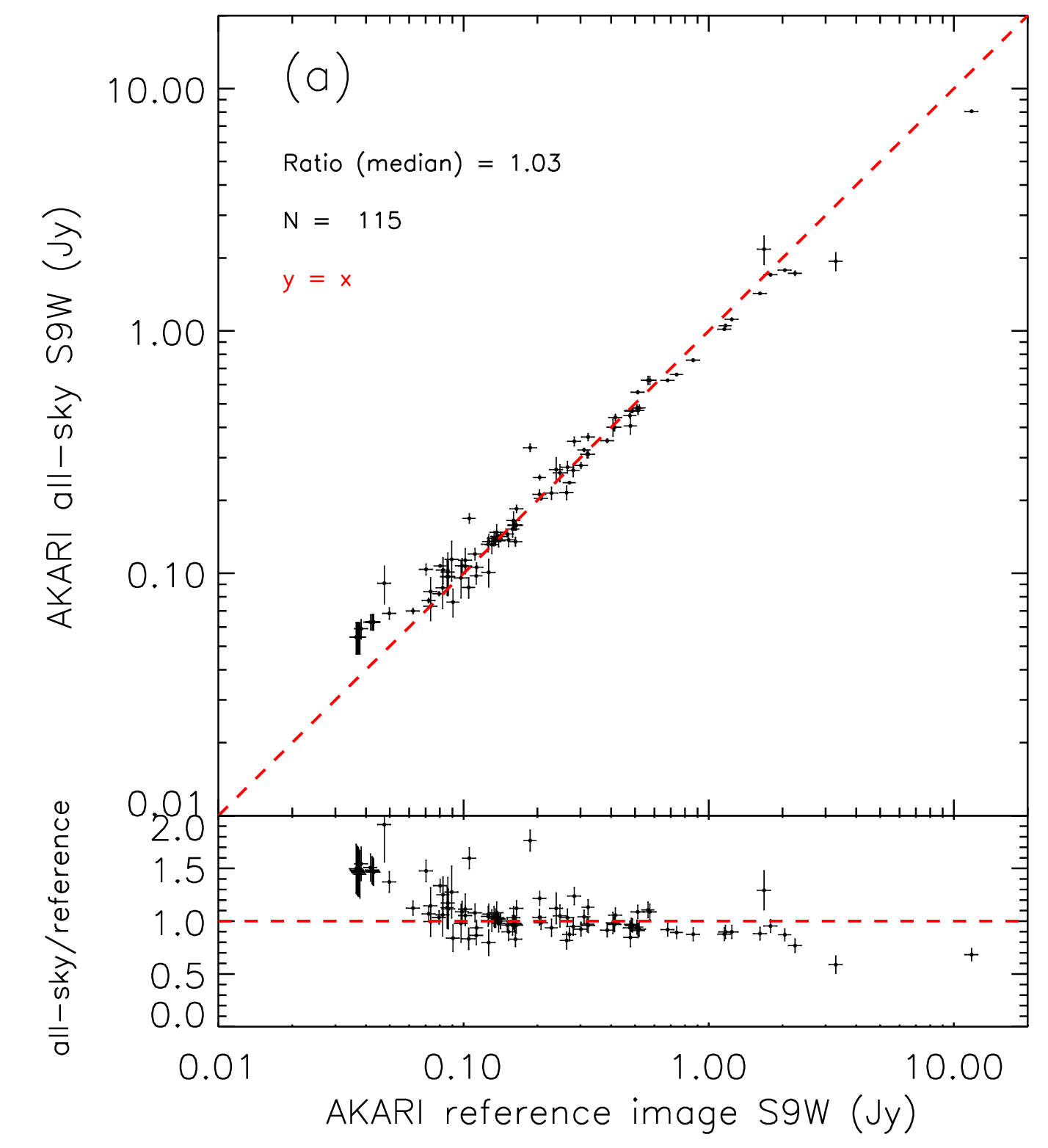} 
  \includegraphics[width=0.48\textwidth]{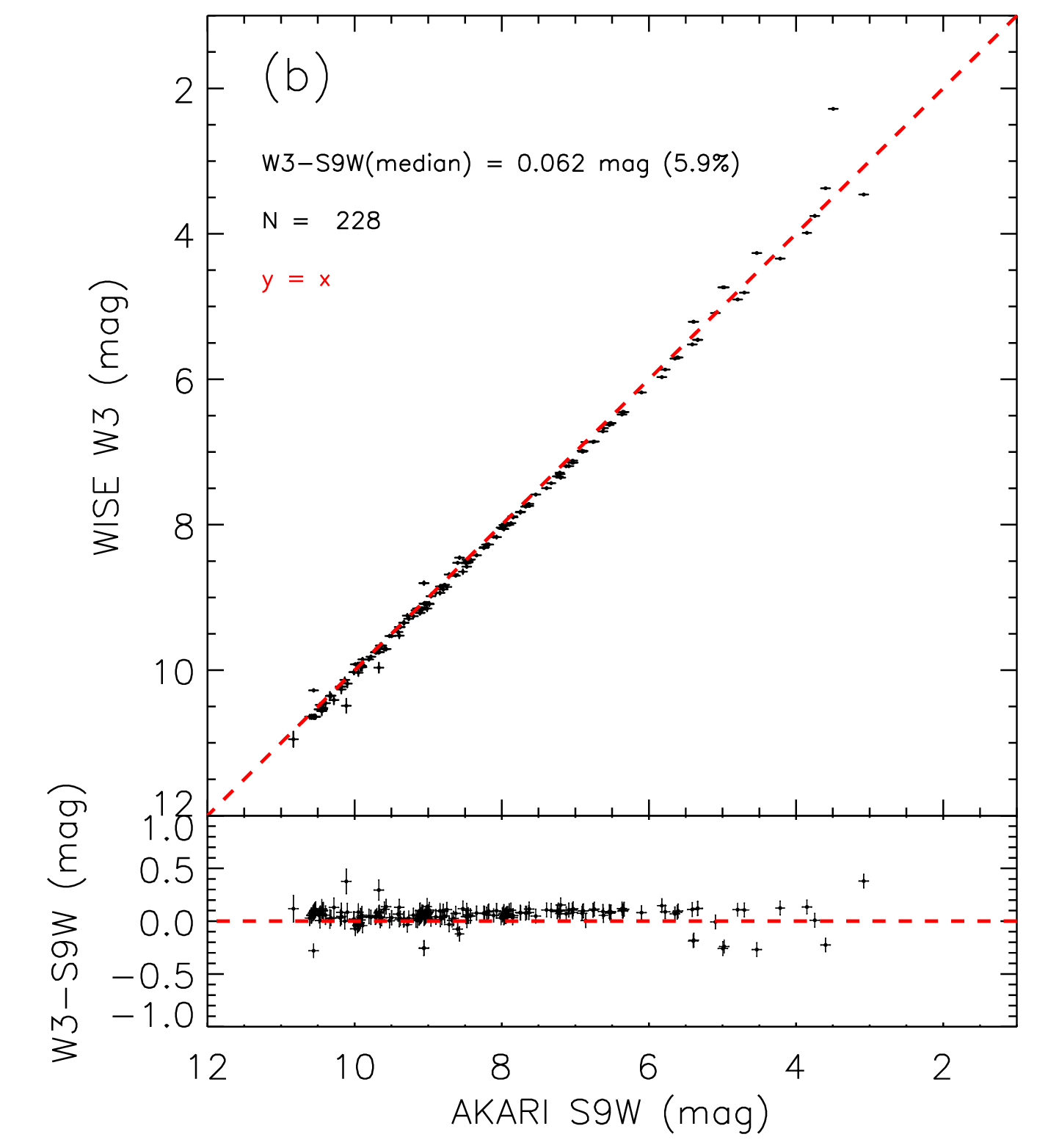} 
 \end{center}
\caption{Comparison of fluxes estimated from the present AKARI reference image 9~$\micron$ and (a) AKARI all-sky survey 9~$\micron$ and (b) WISE W3 12~$\micron$ for the objects yielding grade-S spectra in the present study. Upper parts are the direct comparison of fluxes, and lower parts plot the variations in the fractional fluxes. Only the objects identified as stars after cross-matching with SIMBAD are used in the WISE W3 comparison. Numbers at upper left state the difference between the two bands and the number of objects used in the analysis. The red line indicates $y=x$.}\label{flux_calibration_GradeS}
\end{figure}

\begin{figure}
 \begin{center}
  \includegraphics[width=0.65\textwidth]{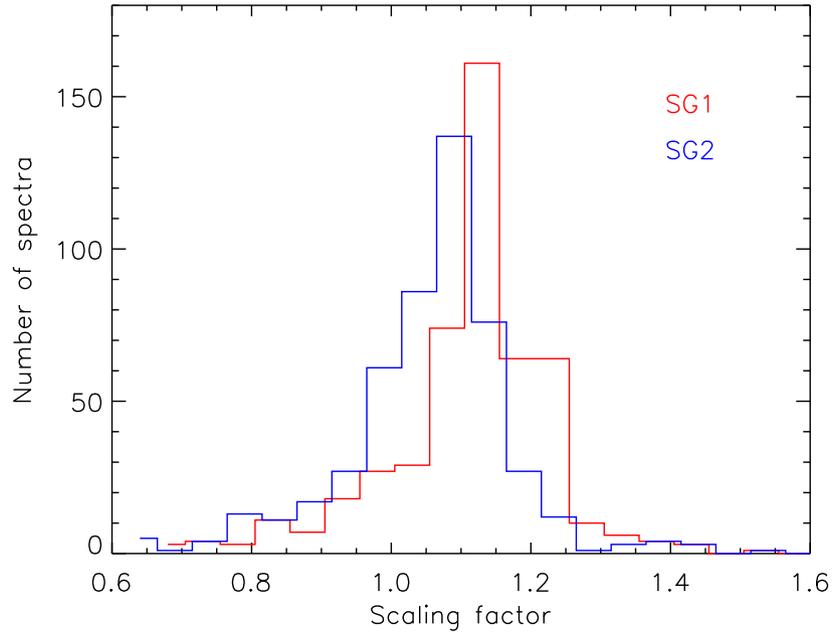} 
 \end{center}
\caption{Histogram of absolute flux calibration factors applied to toolkit-output SG1 and SG2 spectra.}\label{histogram_scaling}
\end{figure}

\begin{figure}
 \begin{center}
  \includegraphics[width=0.48\textwidth]{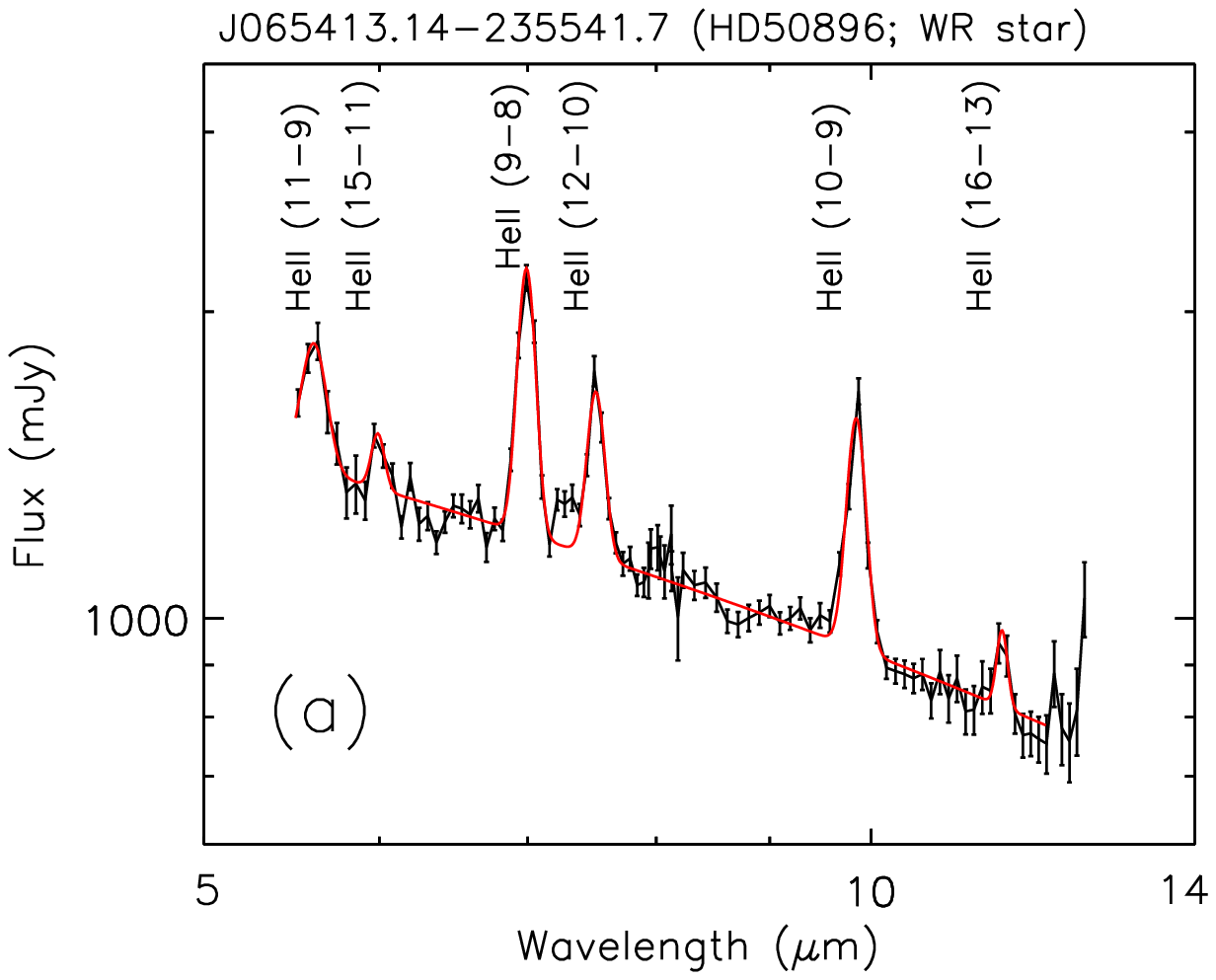}
  \includegraphics[width=0.48\textwidth]{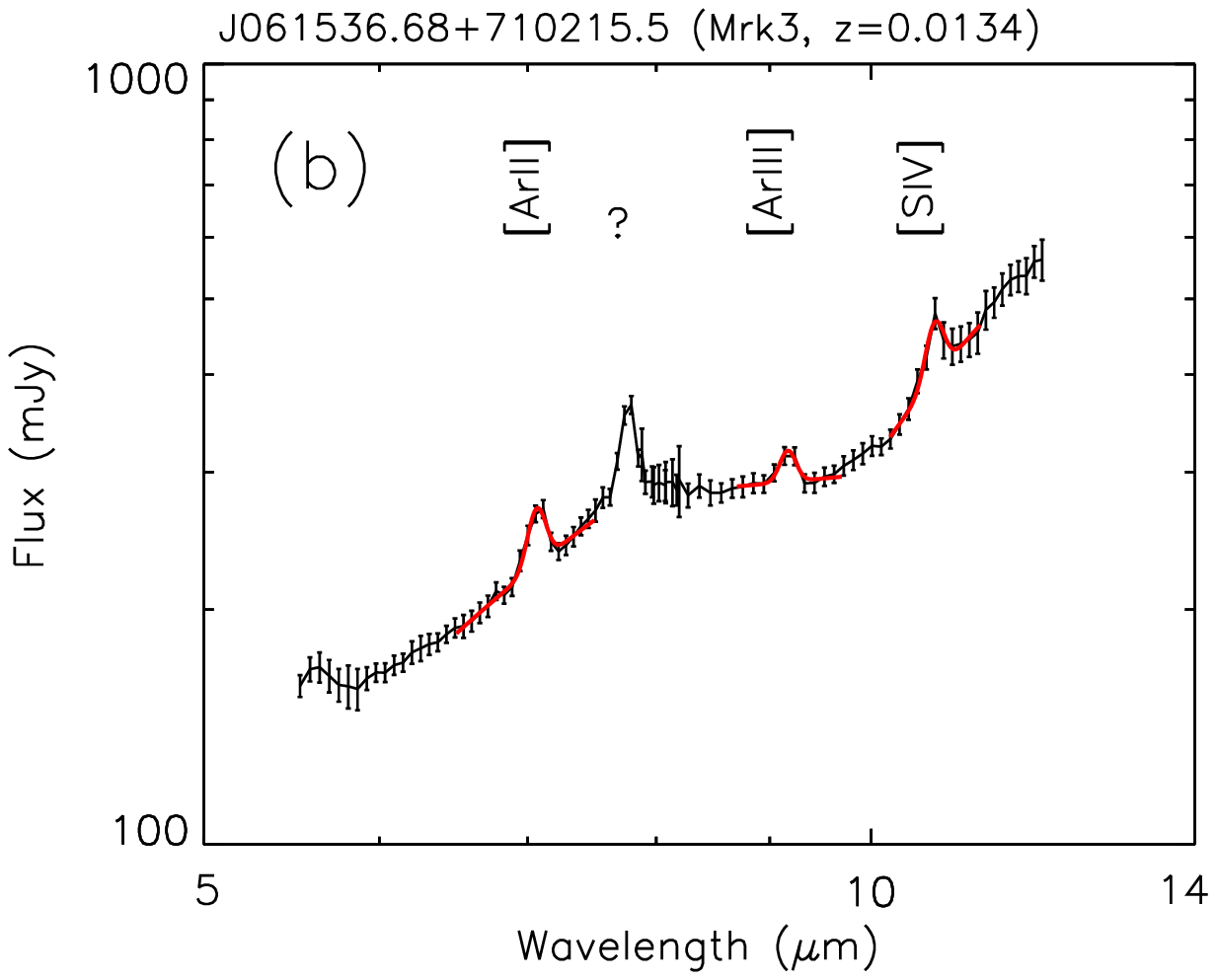}
  \includegraphics[width=0.48\textwidth]{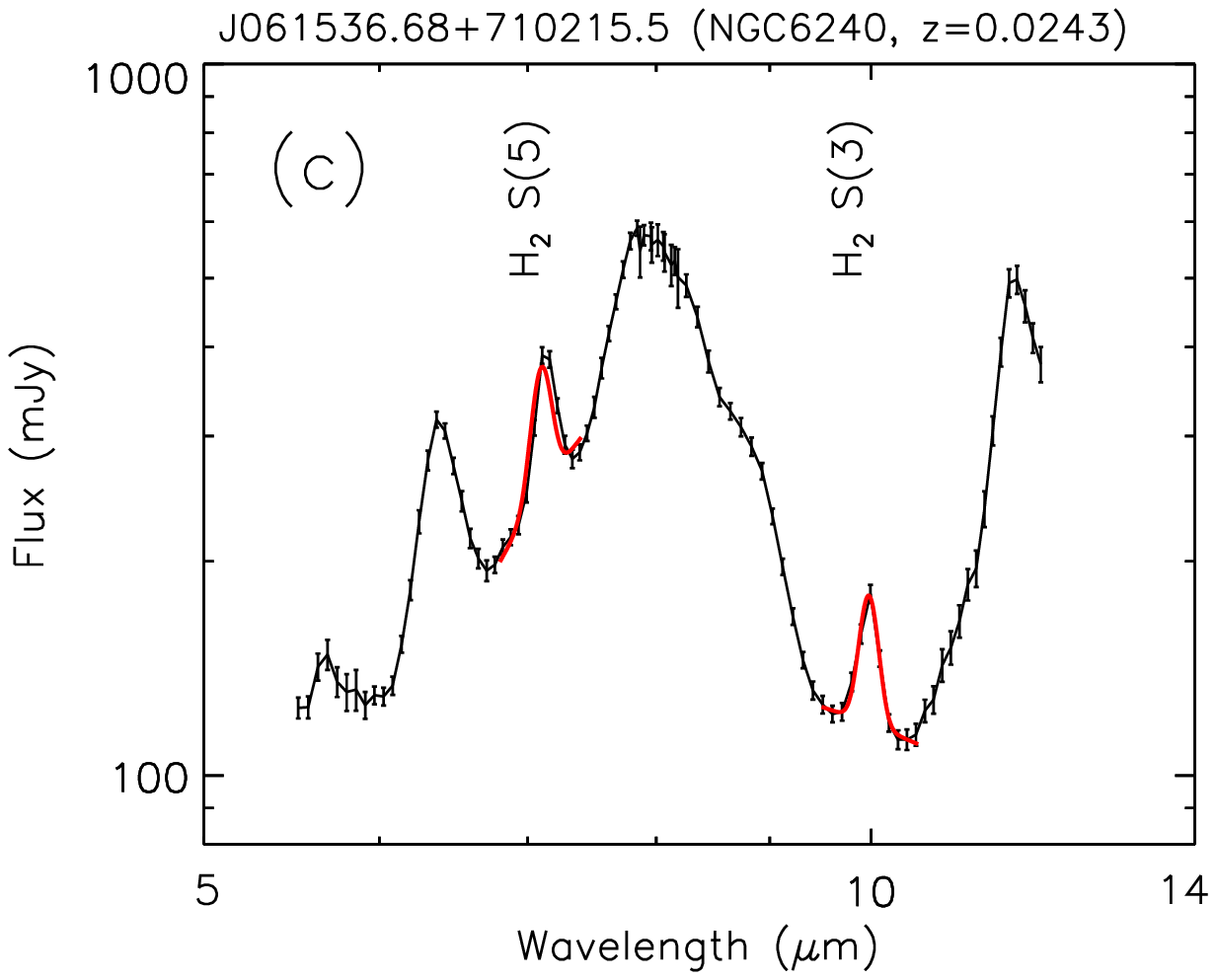}
  \includegraphics[width=0.48\textwidth]{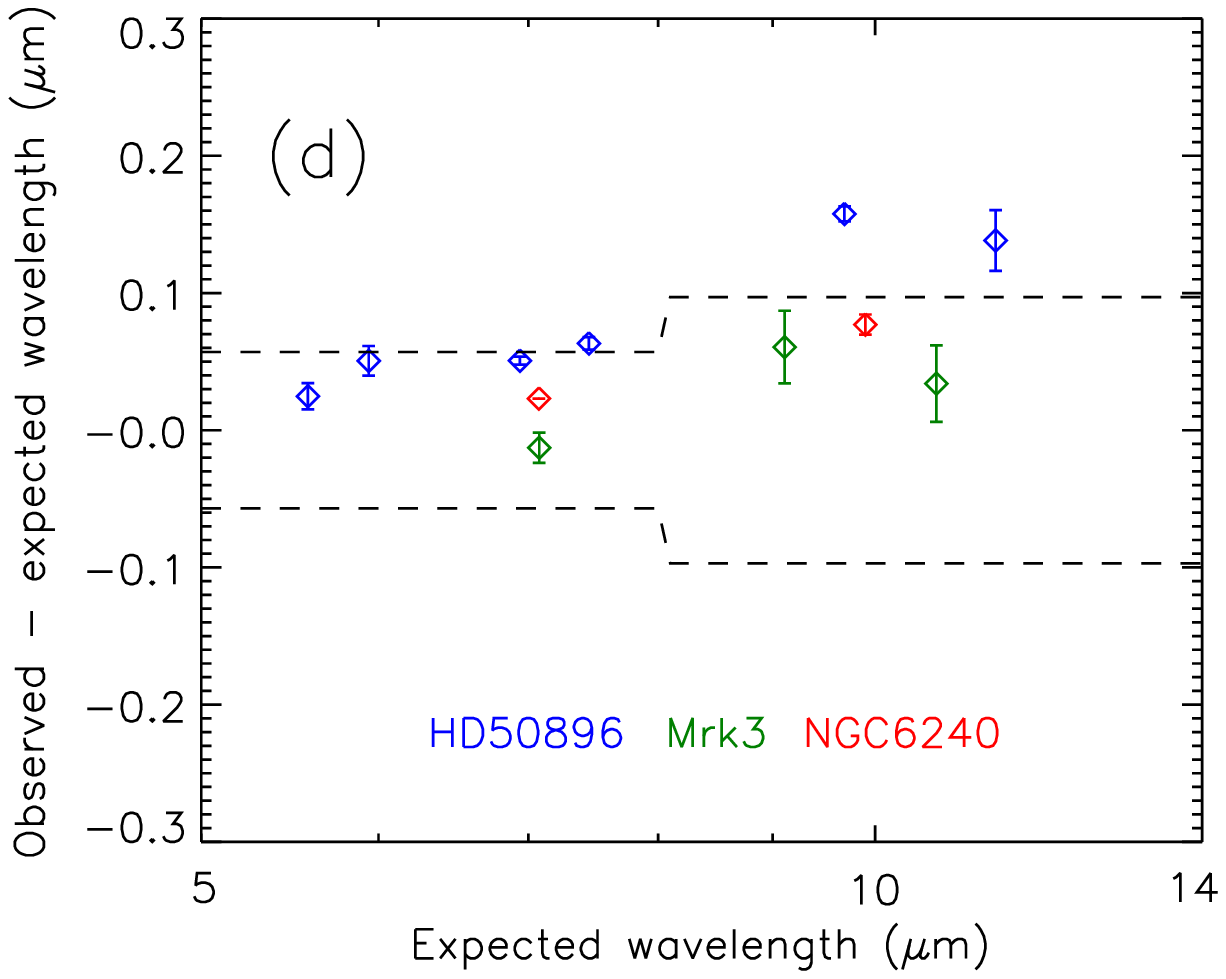}
 \end{center}
\caption{Verification of the wavelength accuracy. (a--c) Spectra with their emission lines and the best-fit results for checking the central wavelength (red curves). The object ID and name are shown along the top of each panel. All emission lines were reproduced by a Gaussian function. The continuum emission was reproduced by a power-law function and a constant in (a) or by a first-order polynomial in (b) and (c). (d) Difference between the observed and expected wavelengths. The data points are color-coded by object. The dashed lines indicate the wavelength difference corresponding to $\pm$1~pixel (0.057~$\micron$/pixel for SG1, 0.097~$\micron$/pixel for SG2; \citealt{Ohyama07}).}\label{wavelength_check}
\end{figure}

\begin{figure}
 \begin{center}
  \includegraphics[width=0.35\textwidth]{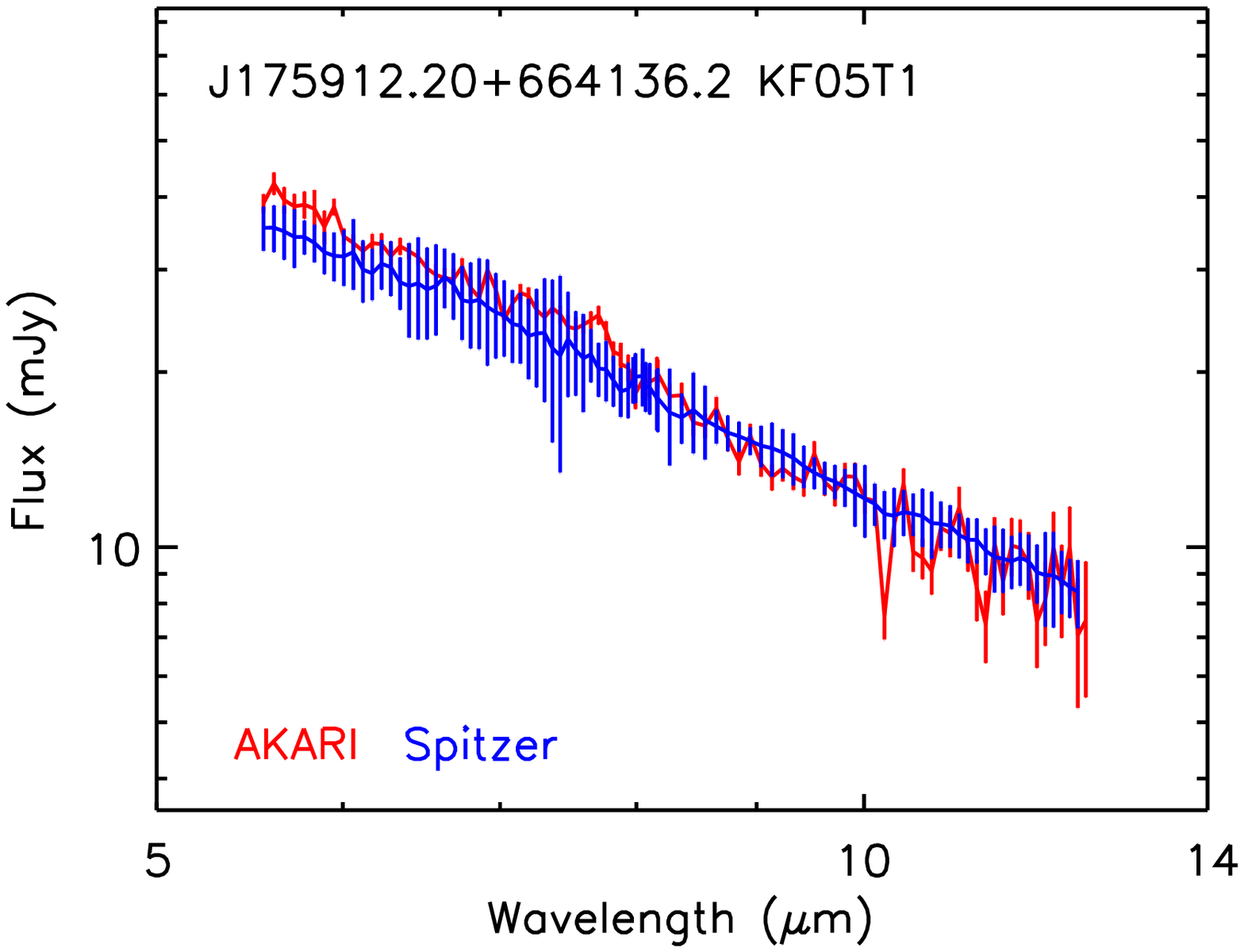}
  \includegraphics[width=0.35\textwidth]{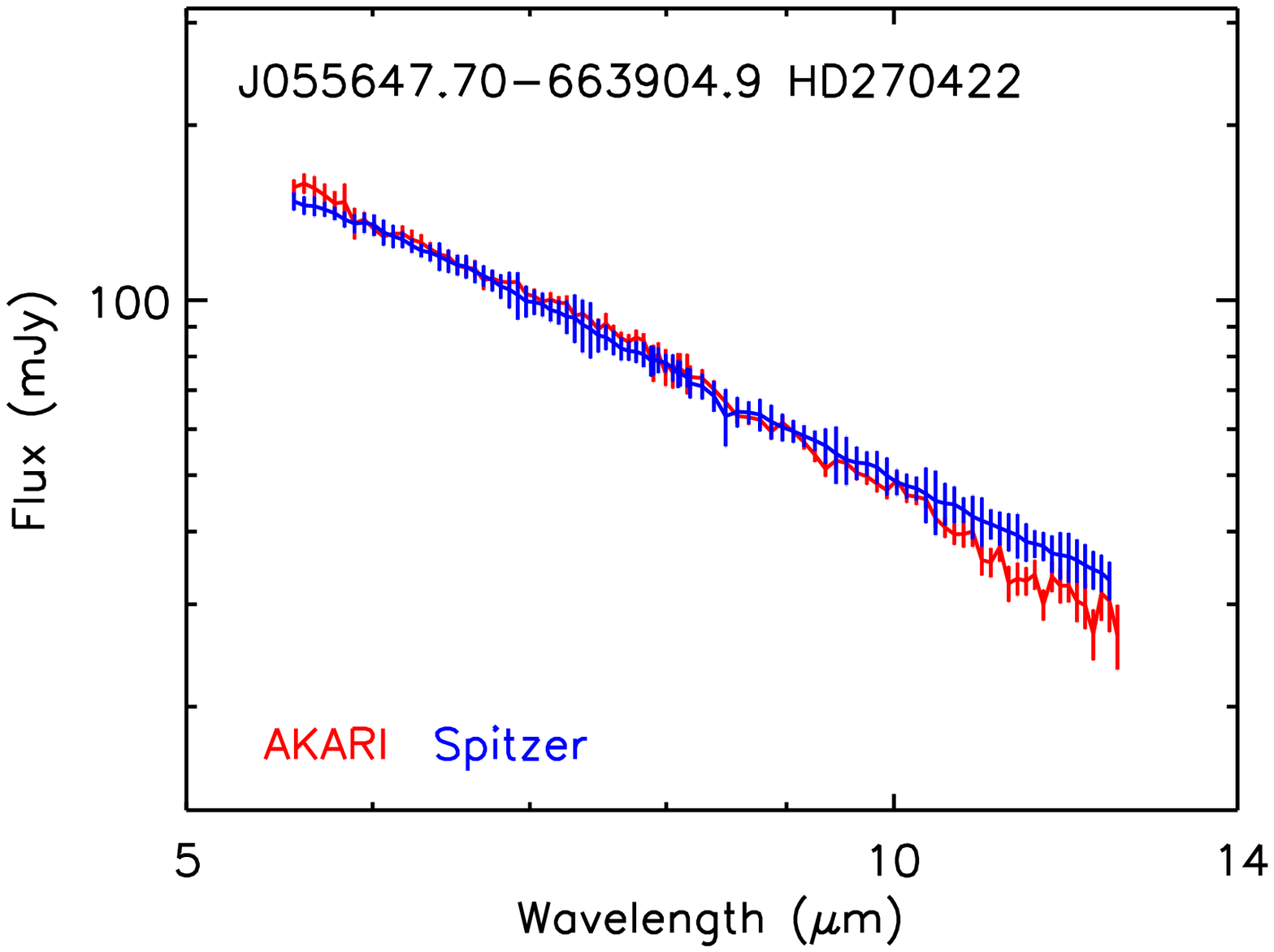}
  \includegraphics[width=0.35\textwidth]{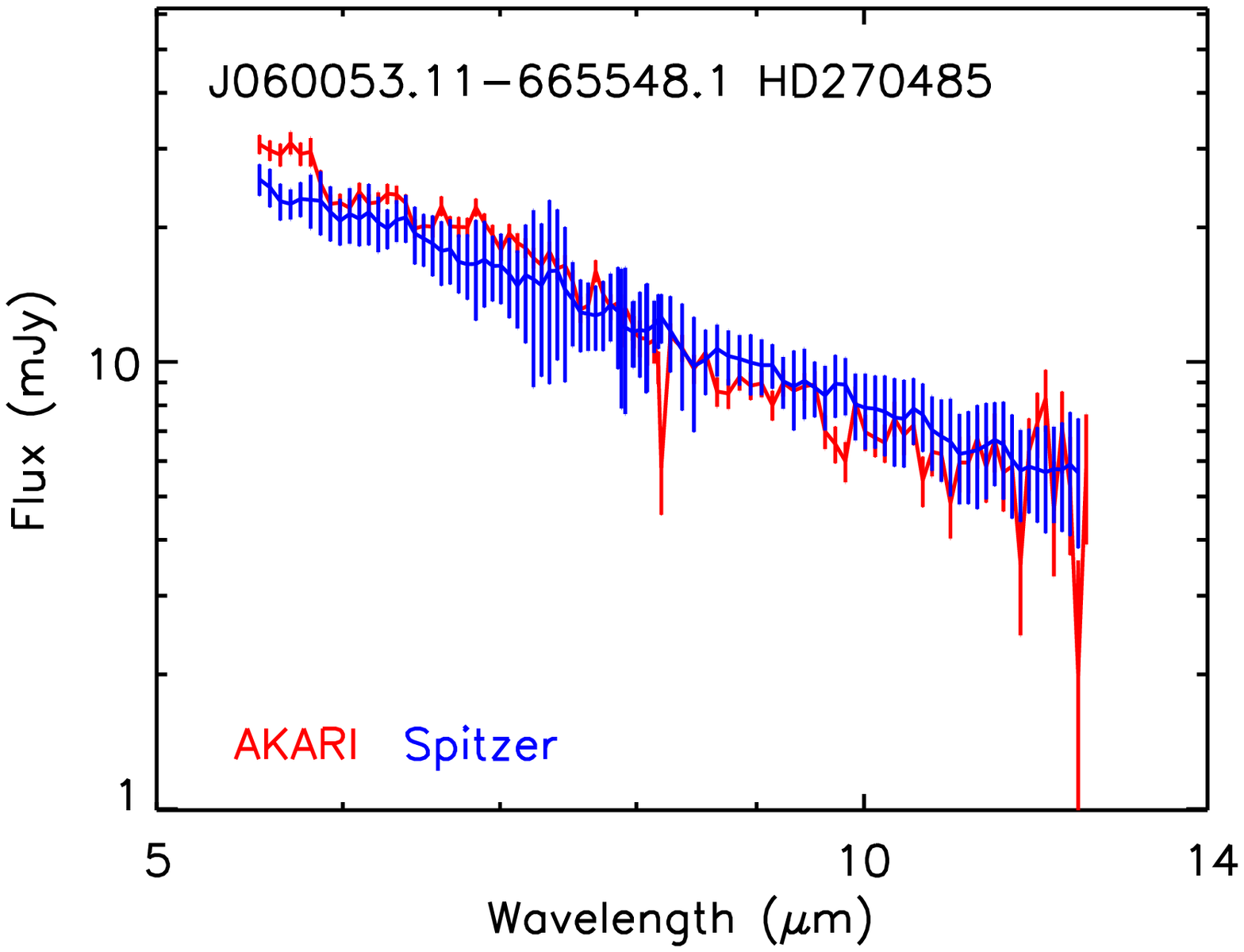}
  \includegraphics[width=0.35\textwidth]{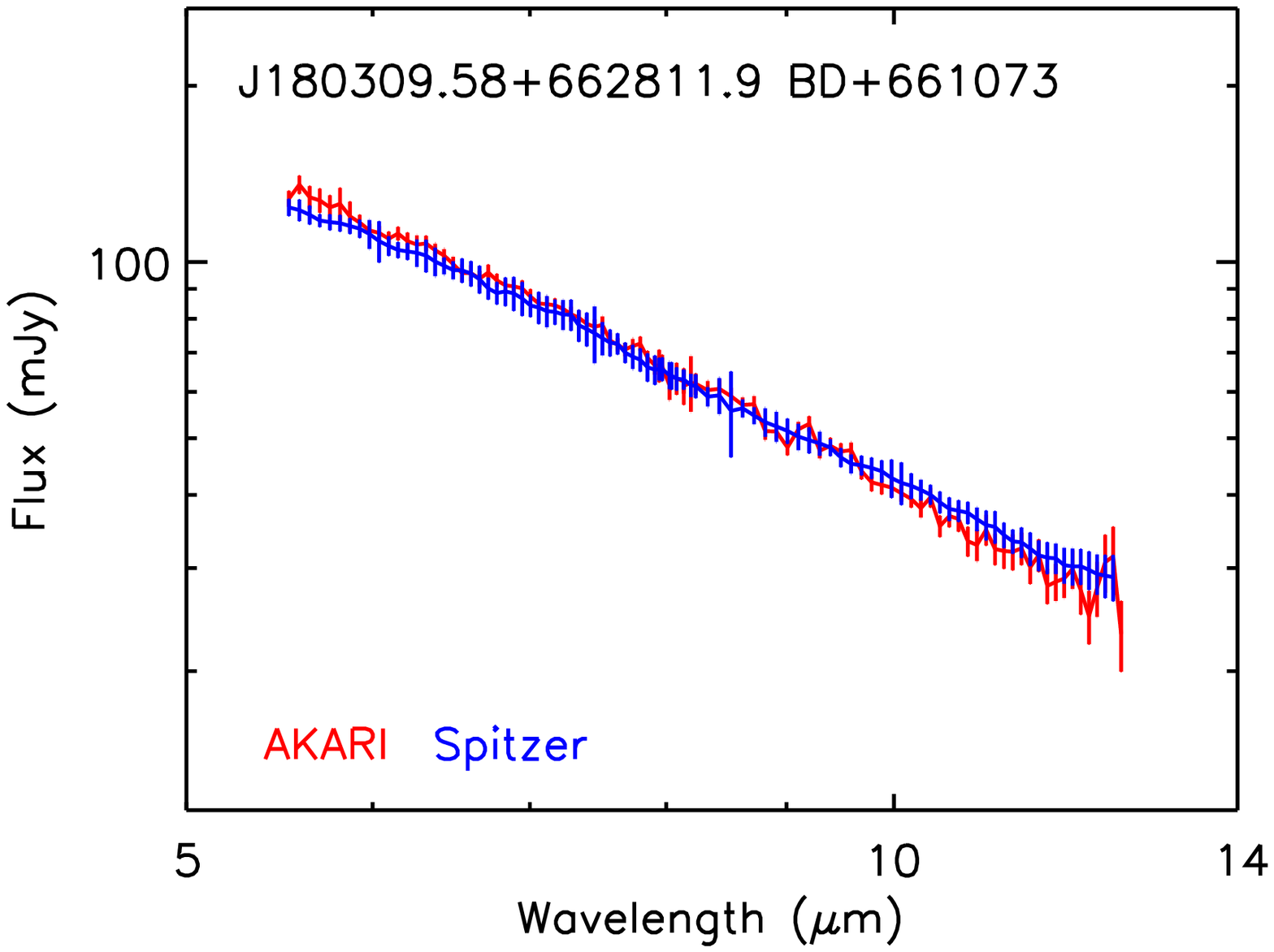}
  \includegraphics[width=0.35\textwidth]{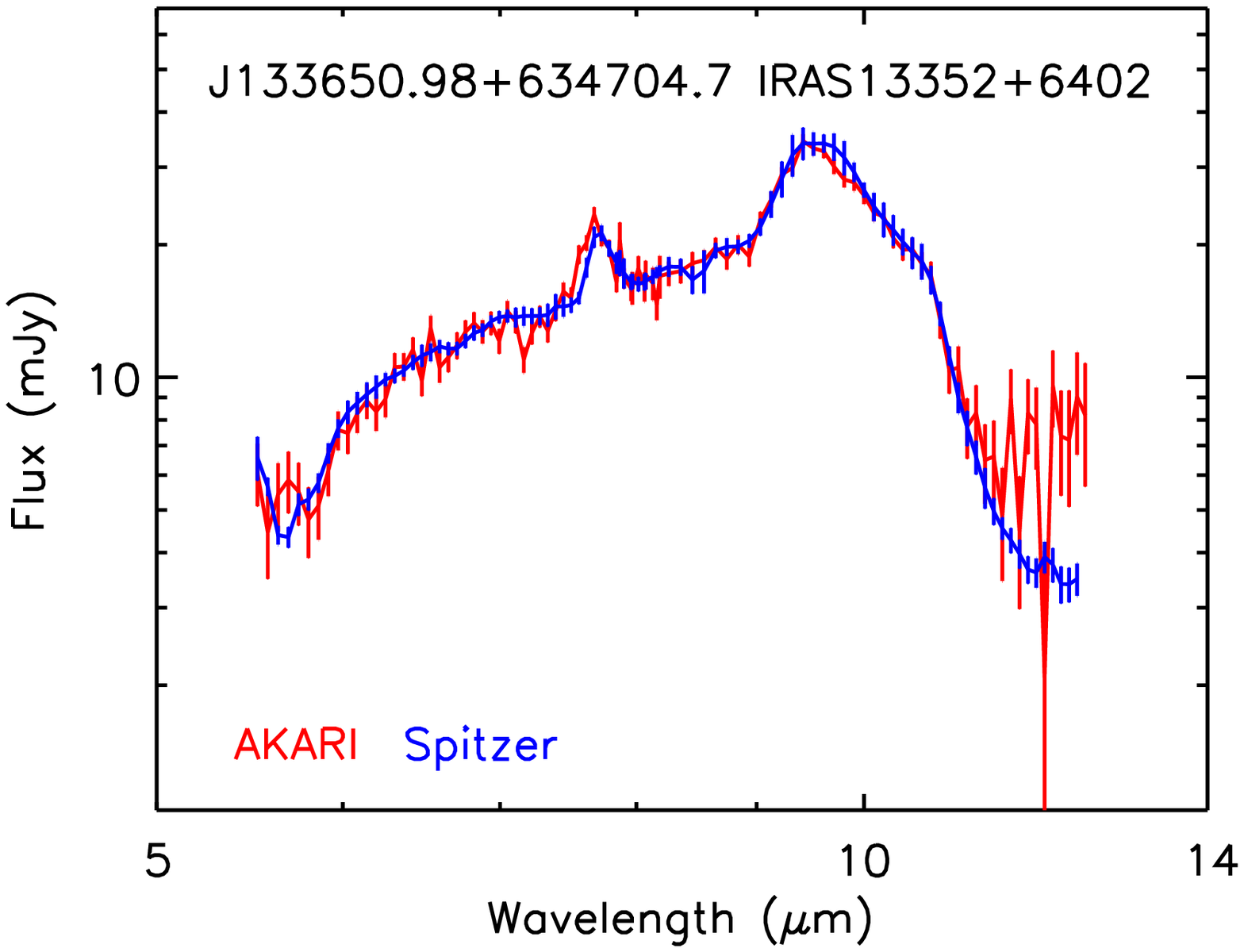}
  \includegraphics[width=0.35\textwidth]{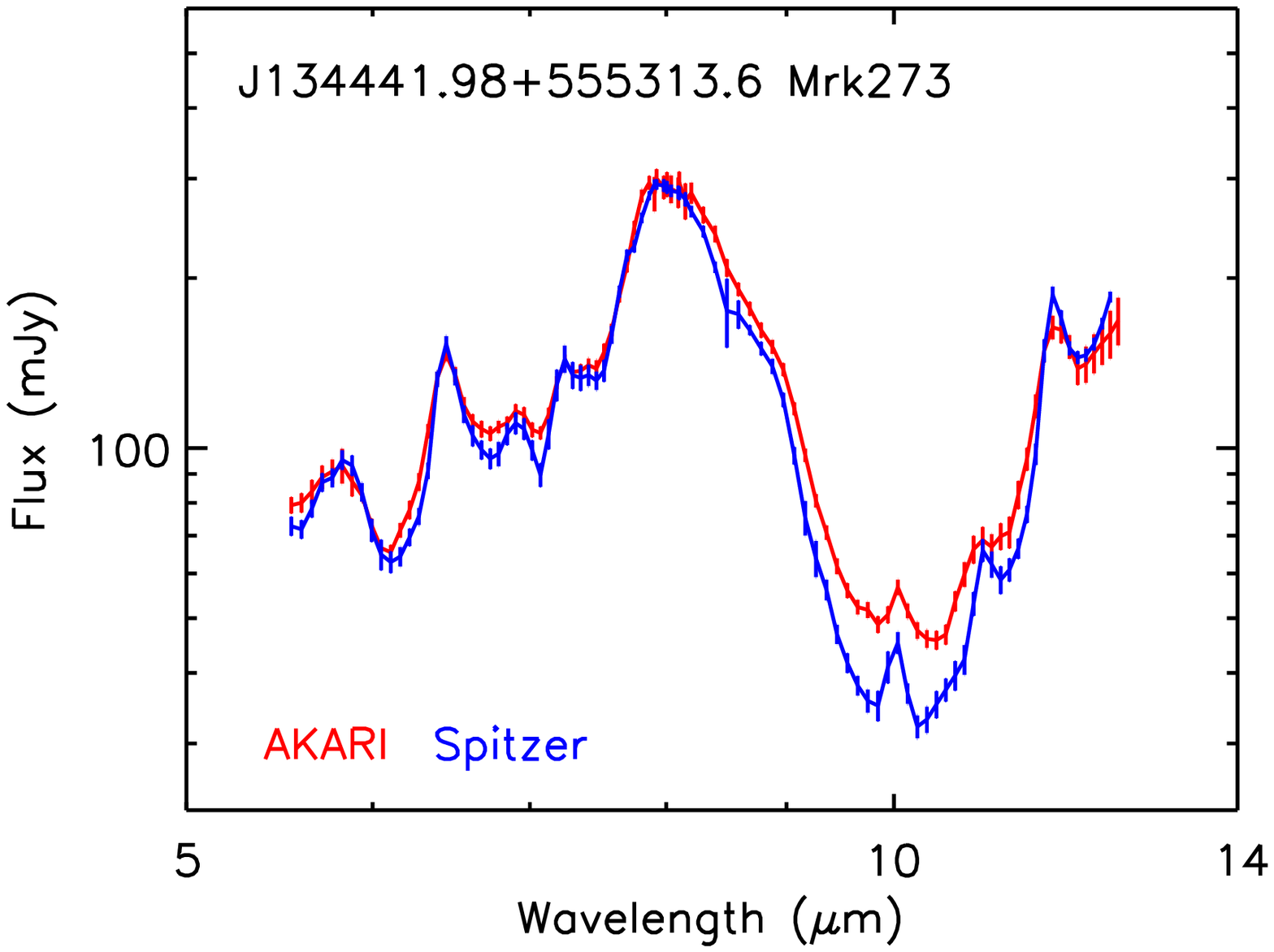}
  \includegraphics[width=0.35\textwidth]{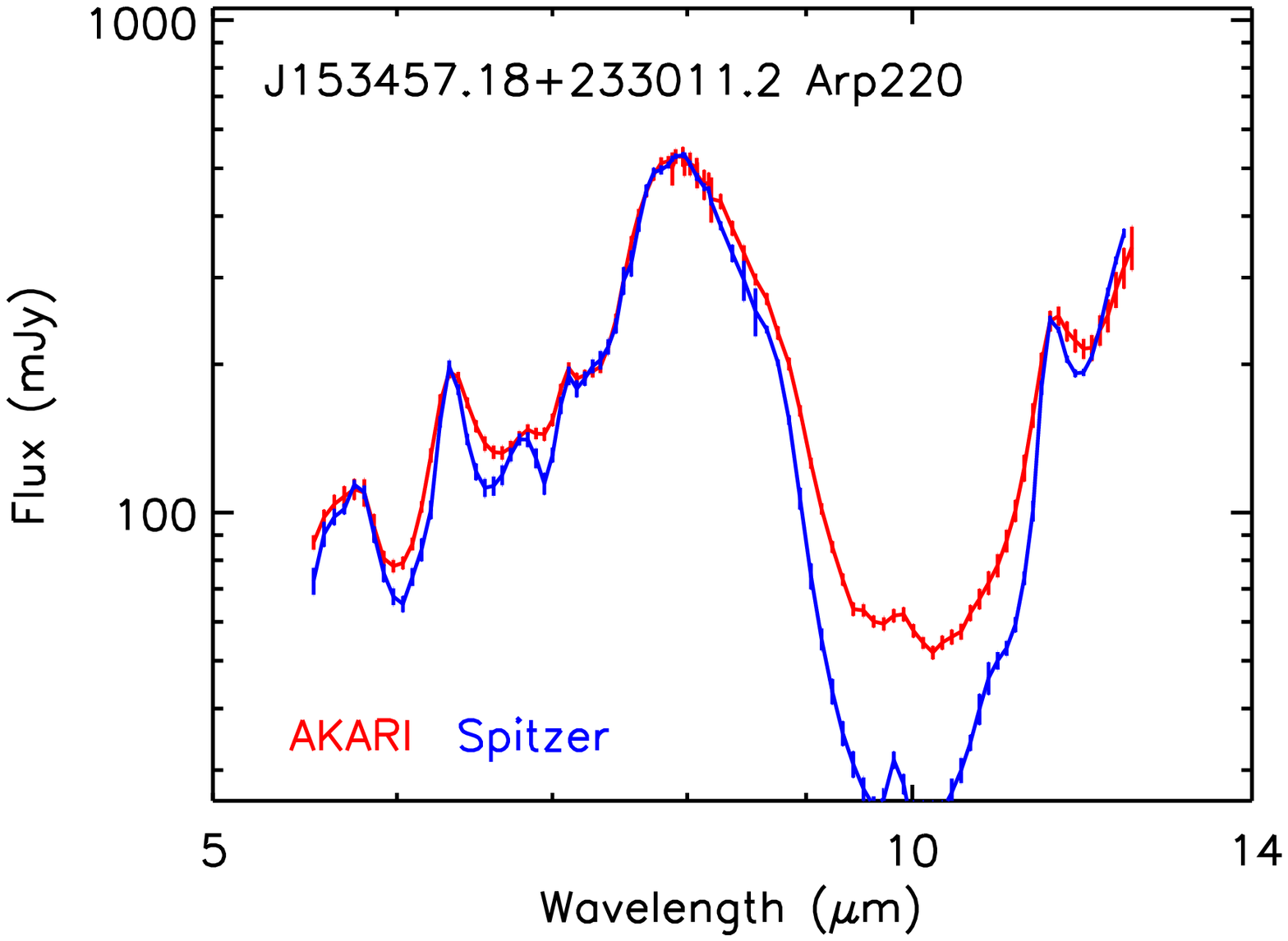}
  \includegraphics[width=0.35\textwidth]{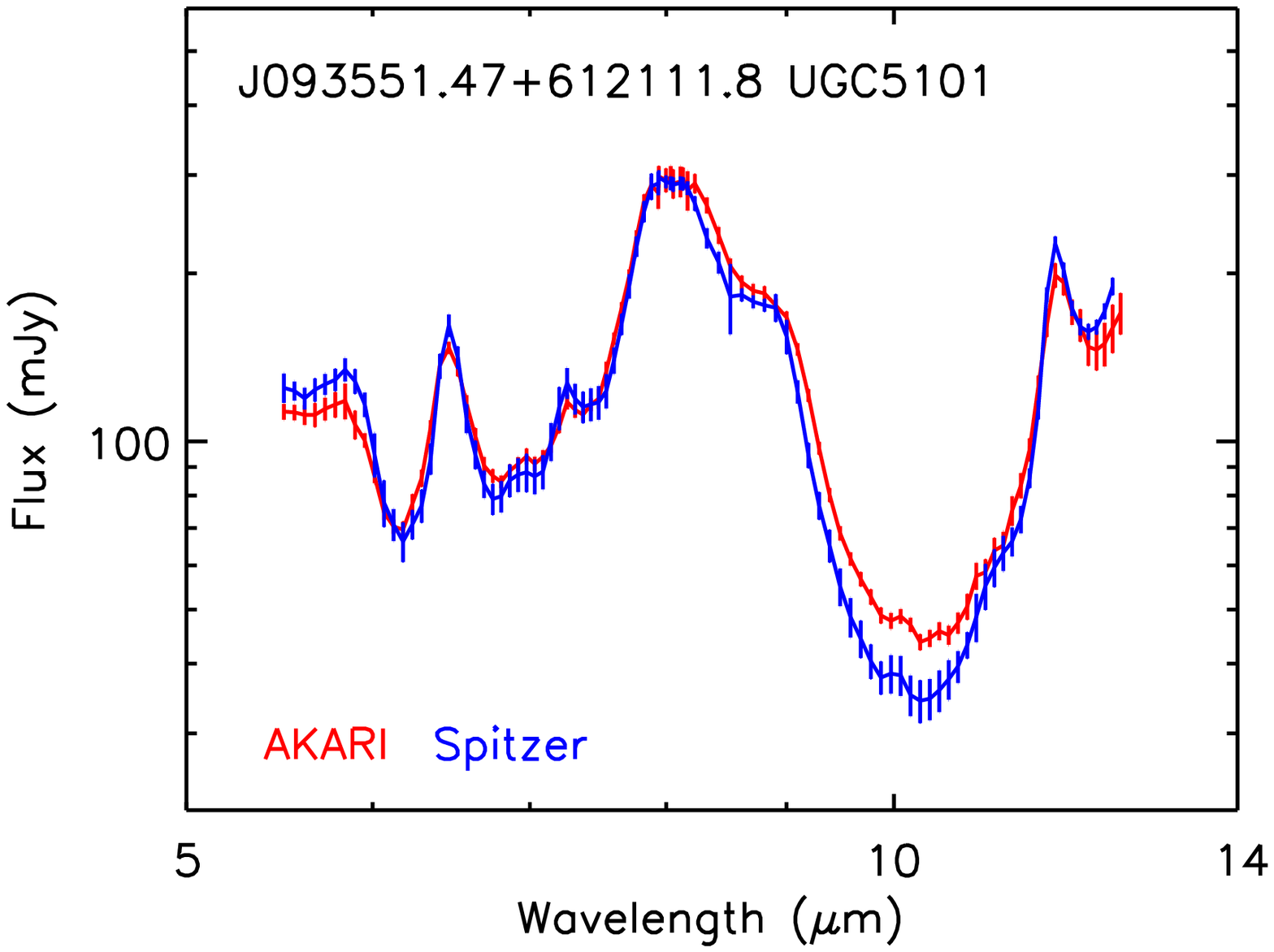}
 \end{center}
\caption{Comparison between AKARI/IRC MIR-S slit-less spectra (red) and Spitzer/IRS SL slit spectra (blue). The first four objects are stars and the latter four are galaxies. Object ID and name are shown along the top of each panel.}\label{spitzer_hikaku}
\end{figure}

\begin{figure}
 \begin{center}
  \includegraphics[width=0.7\textwidth]{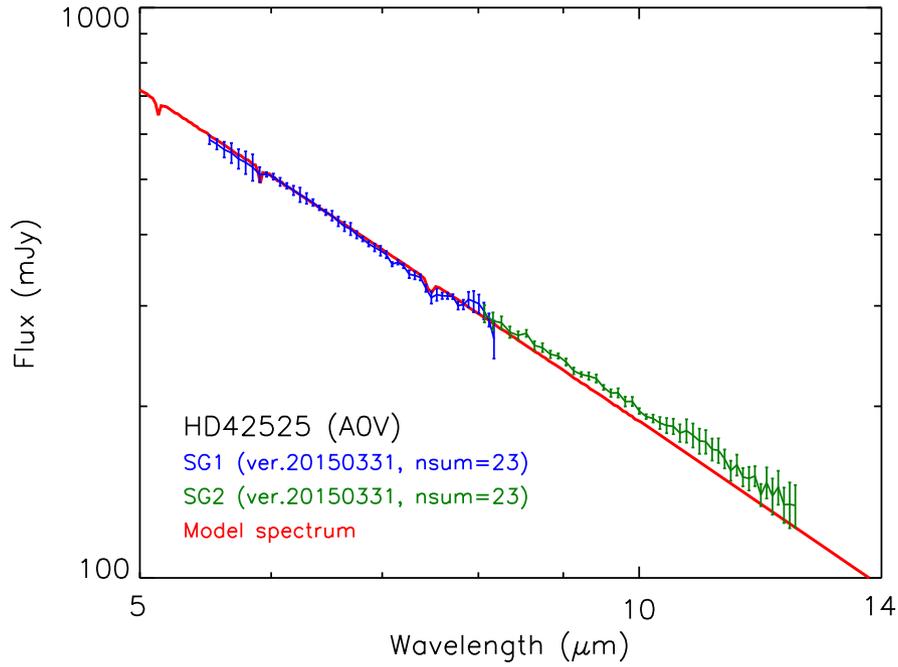}
 \end{center}
\caption{Comparison between the modeled spectrum of HD42525 (red) and the spectrum extracted by the toolkit ver. 20150331 (SG1: blue; SG2: green). Object HD42525 is an A0V type star. The model is given in \citet{Cohen03}.}\label{spec_response_hikaku}
\end{figure}

\begin{figure}
 \begin{center}
  \includegraphics[width=0.7\textwidth]{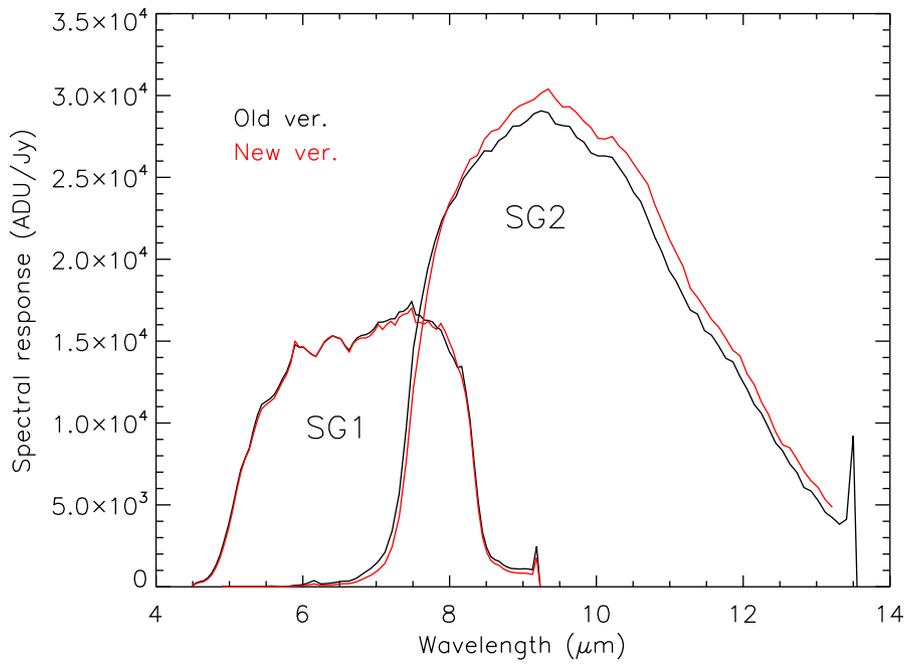} 
 \end{center}
\caption{Comparison between the old (black) and new (red) spectral response curves.}\label{response_hikaku}
\end{figure}

\begin{figure}
 \begin{center}
  \includegraphics[width=0.48\textwidth]{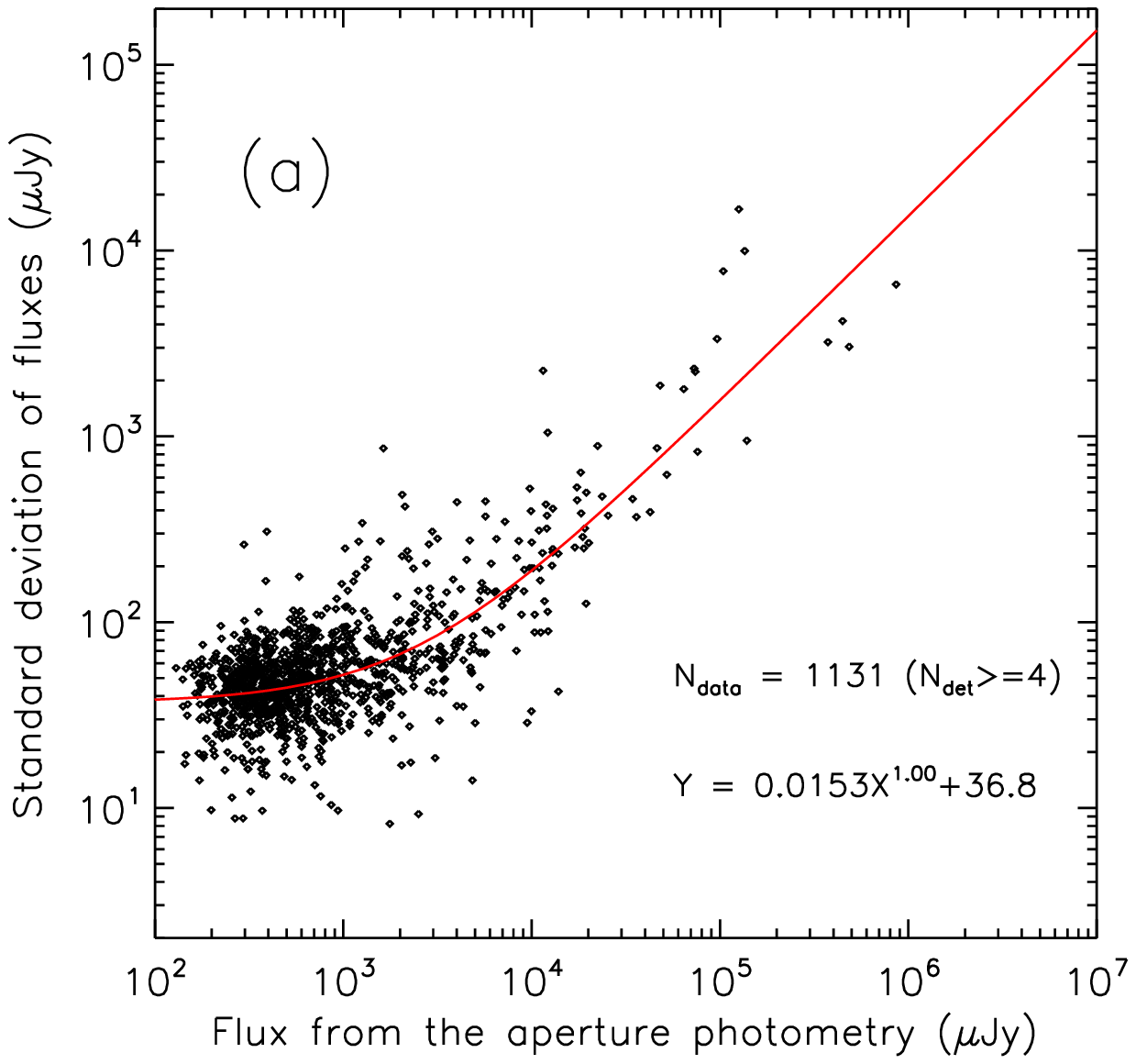} 
  \includegraphics[width=0.48\textwidth]{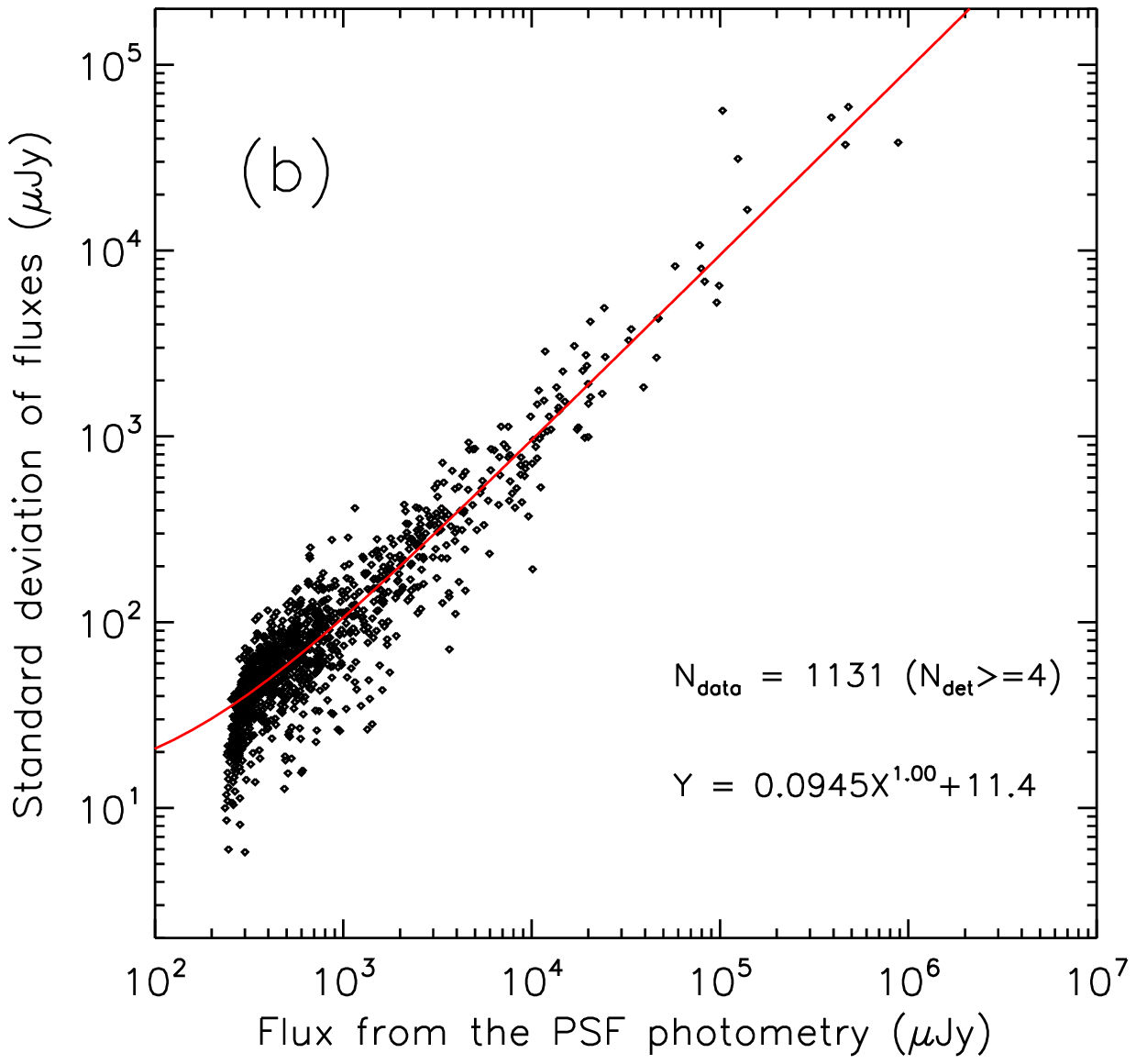} 
 \end{center}
\caption{Standard deviations of the 9~$\micron$ flux as functions of average 9~$\micron$ flux for the point sources detected four times or more. Fluxes were measured from (a) the aperture photometry and (b) the PSF photometry. Red curves indicate the best-fit relations for estimating the random error. The numbers at bottom right are the number of data points used in this analysis and the parameters of the best-fit relations.}\label{randomerror_suitei}
\end{figure}

\begin{figure}
 \begin{center}
  \includegraphics[width=0.48\textwidth]{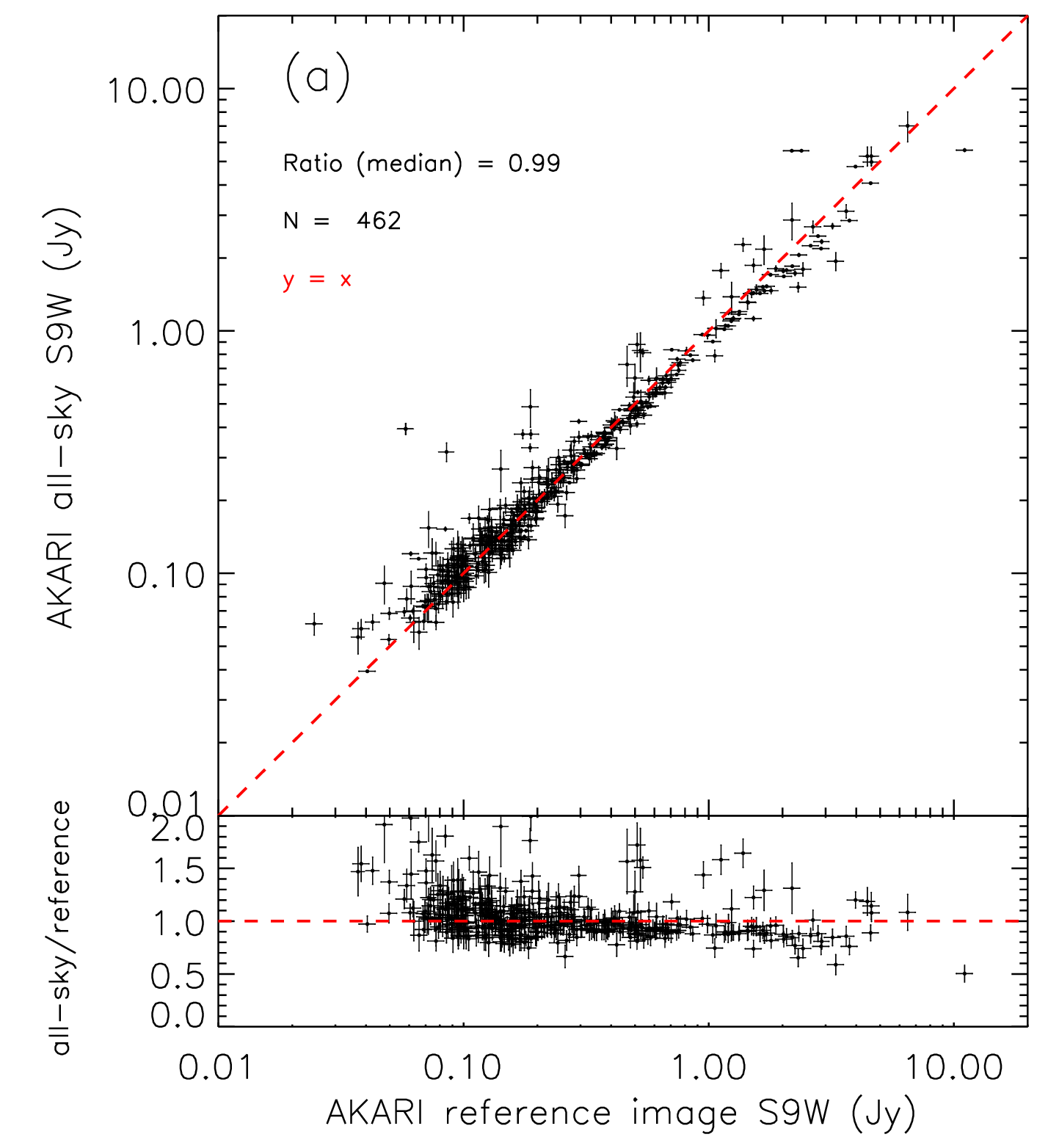} 
  \includegraphics[width=0.48\textwidth]{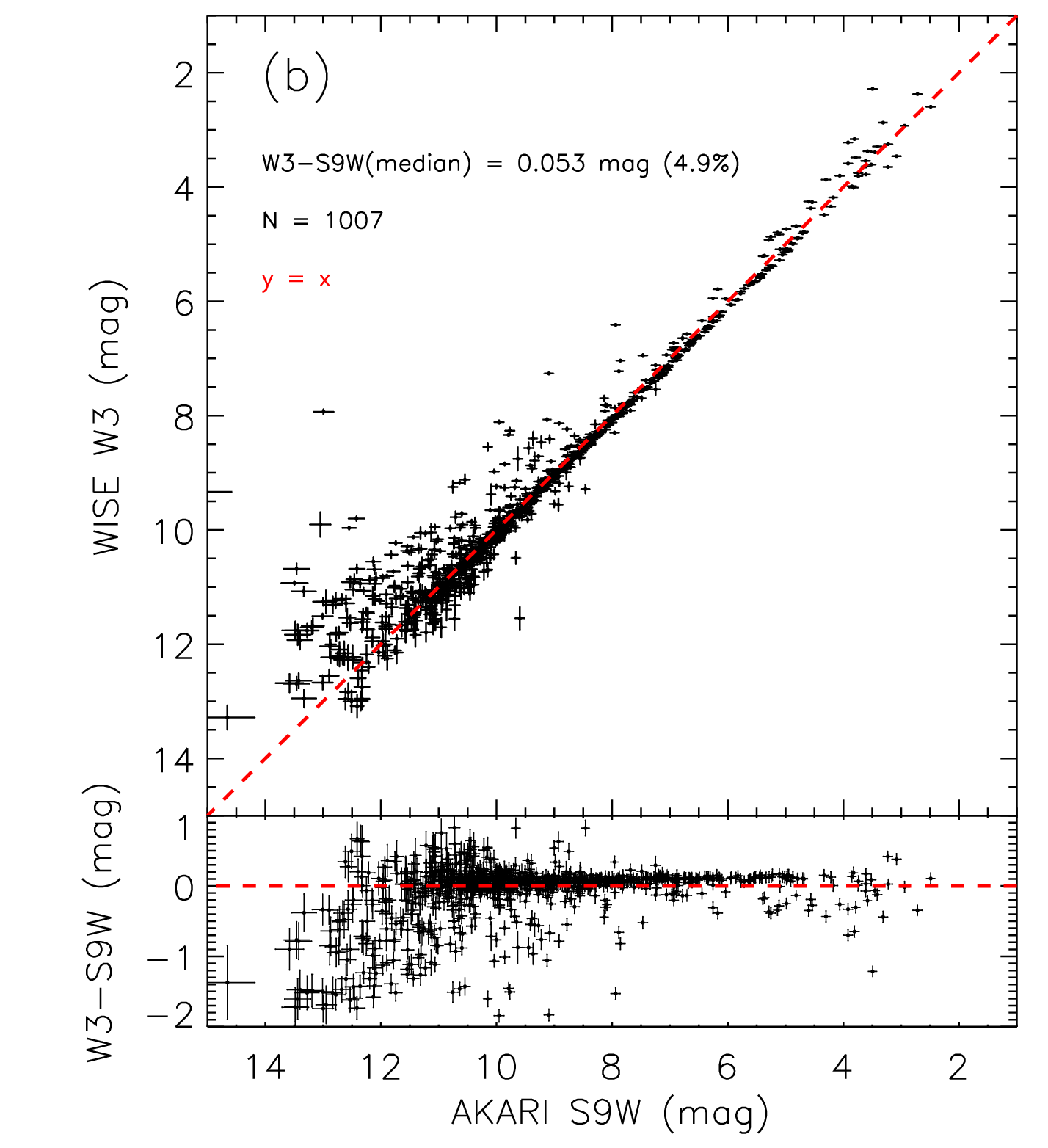} 
 \end{center}
\caption{Comparison of fluxes between the present AKARI reference image 9~$\micron$ and (a) AKARI all-sky survey 9~$\micron$, (b) WISE W3 12~$\micron$, for all sources in the present 9~$\micron$ point-source catalogue. Upper parts show the direct flux comparisons, and lower parts indicate the variations in the fractional fluxes. Only objects identified as stars after cross-matching with SIMBAD are used in the comparison with WISE W3. Numbers at upper left indicate the difference between the two bands and the number of objects used in the analysis. The red line indicates $y=x$.}\label{flux_calibration_all}
\end{figure}

\begin{figure}
 \begin{center}
  \includegraphics[width=0.48\textwidth]{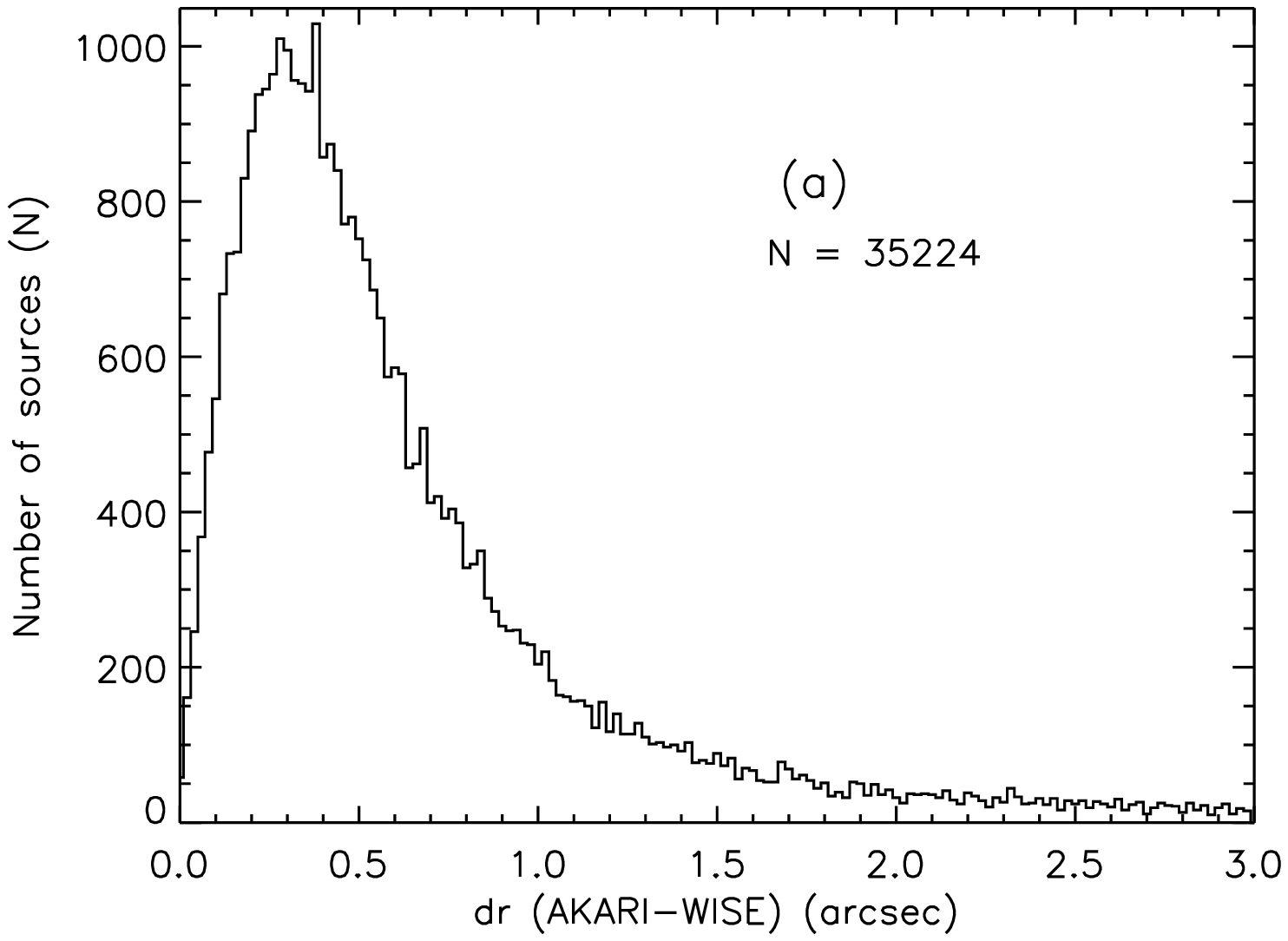} 
  \includegraphics[width=0.48\textwidth]{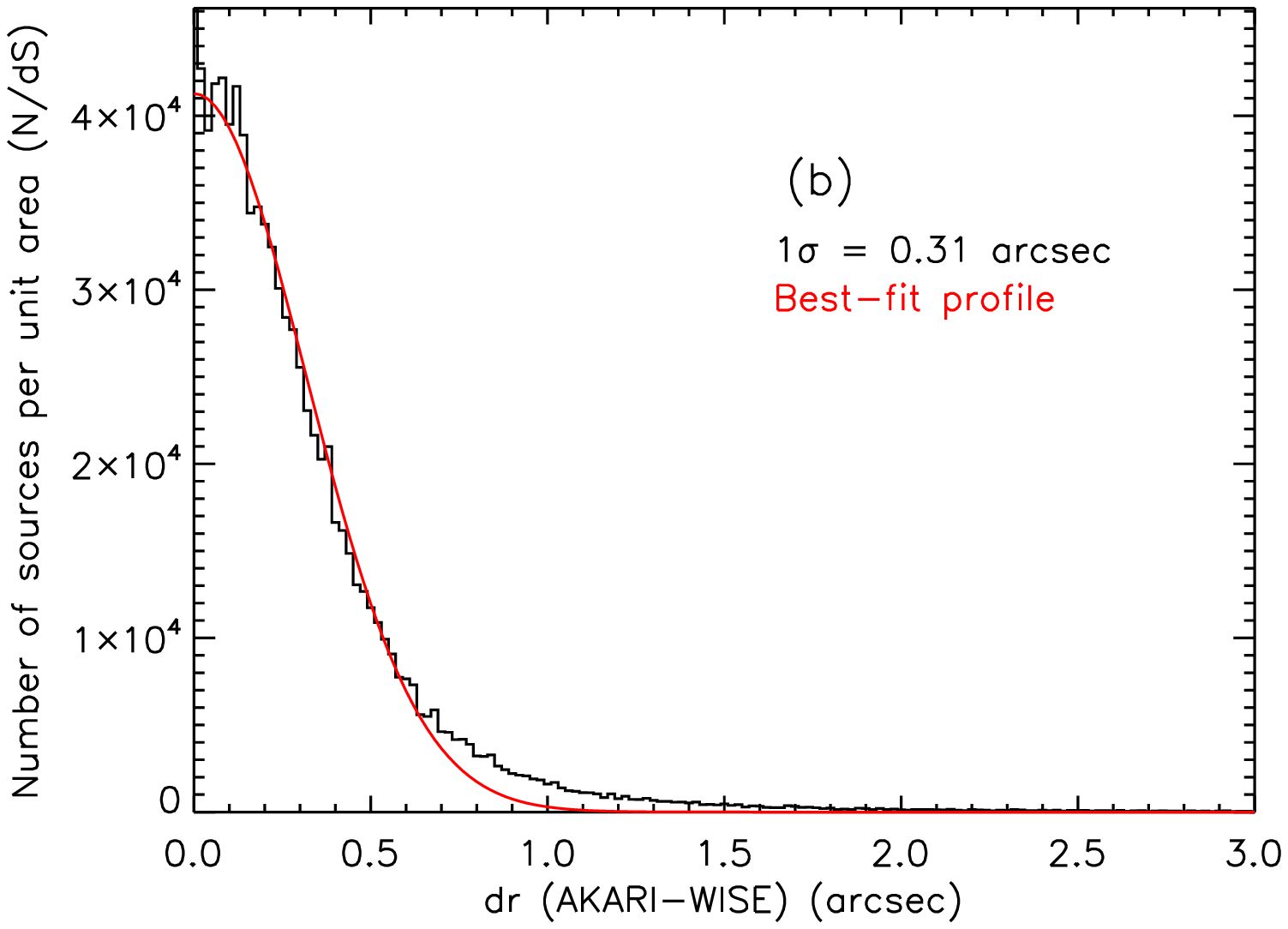} 
 \end{center}
\caption{(a) Histogram of the positional difference between the present 9~$\micron$ point-source catalogue and the WISE all-sky source catalogue. (b) Same histogram as (a), but with the vertical axis divided by the area of the circular ring, and the best-fit Gaussian function (red curve). The upper right numbers indicate the total number of objects used in the analysis in (a), and the best-fit standard deviation in (b).}\label{position}
\end{figure}

\begin{figure}
 \begin{center}
  \includegraphics[width=0.7\textwidth]{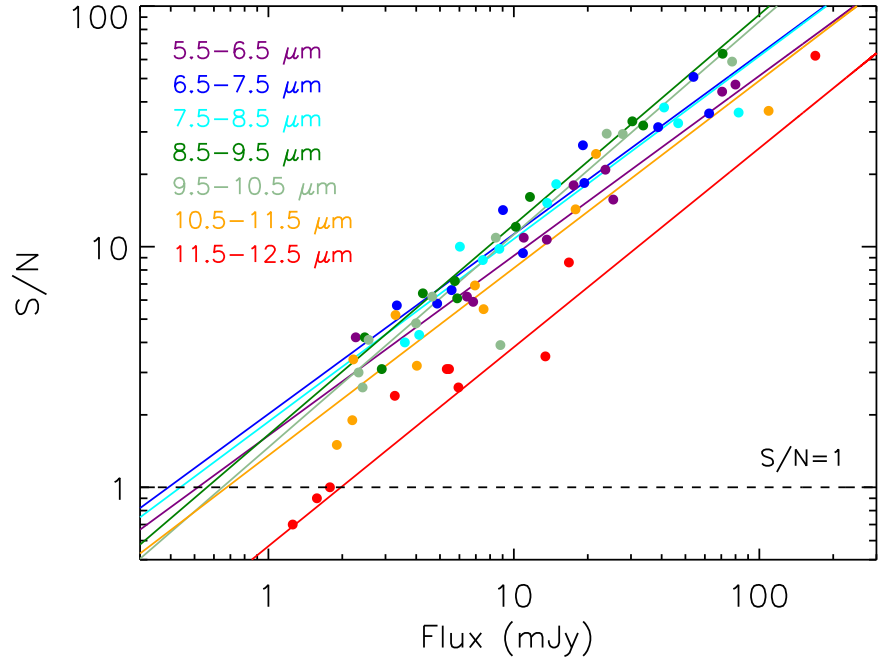} 
 \end{center}
\caption{S/N--flux relations of the spectra. The median S/Ns versus flux in each wavelength range are plotted for the ten spectra in Figure~\ref{stability}.  The lines are the best fit power-law relations in the seven wavelength ranges. The dashed line corresponds to S/N=1.}\label{stability_dependence}
\end{figure}

{\appendix 

\section{Updates of the spectroscopic toolkit for the MIR-S spectroscopy}

Figure~\ref{spec_response_hikaku} shows the observed and modeled spectra of HD~42525 analyzed by the spectroscopic toolkit currently available on the web\footnote{https://www.ir.isas.jaxa.jp/AKARI/Observation/support/IRC/} (ver. 20150331).
To estimate the total flux, the spectrum was extracted from a 23-pix wide rectangular region, and the sky level was determined in a 6-pix wide contiguous rectangular region.
HD~42525 is a standard source for the spectroscopic flux calibration of AKARI/IRC (\citealt{Ohyama07}), so the modeled and observed spectra should be well matched.
We, however, found systematic differences between the observed and modeled spectra in the SG2 range.
Since the spectroscopic toolkit used in \citet{Ohyama07} to perform the spectroscopic calibration is different from the latest version employed in the present work, we updated the calibration files to guarantee the flux level of the present spectroscopic catalogue.
The flux calibration depends on the data reduction procedure, how the spectrum is extracted and how the sky background is subtracted.
Therefore, the same procedure must be used both for calibration stars and observations.
We attributed the difference from the previous calibration to the spectral response curves to provide consistent results.
Figure~\ref{response_hikaku} compares the old and new spectral response curves.
As expected from Figure~\ref{spec_response_hikaku}, the new spectral response curve of SG2 is 10~\% larger than the old curve, whereas the old and new spectral response curves of SG1 are roughly the same.
During the verification, the current aperture correction in the spectroscopic toolkit was found to be inadequate, because it ignored the wavelength dependency.
A fraction of the flux missed by the small aperture is wavelength-dependent, especially in the SG2 range.
Therefore, our updated the aperture correction in the spectroscopic toolkit also accounts for the wavelength dependency.
The data in the present study were reduced using the updated spectroscopic toolkit, which is available with the spectral and point-source catalogues.

\section{Details of the point source catalogue}

\subsection{Evaluation of the flux uncertainty}

The present 9~$\micron$ point source catalogue includes two types of flux uncertainty: the systematic calibration error of 6\% (\citealt{Tanabe08}) and a random error estimated from the flux repeatability of sources detected four times or more.
Figure~\ref{randomerror_suitei} shows the standard deviations of the 9~$\micron$ fluxes as functions of the average 9~$\micron$ flux of sources detected four times or more.
Fitting a power-law function ($y=ax^b+c$) to the data, we obtained the relation between the flux and its standard deviation ($F_{\mathrm{best}}$).
The random error in the fluxes of all sources was then estimated as $F_{\mathrm{best}}/\sqrt{N_\mathrm{det}}$, where $N_{\mathrm{det}}$ is the number of detections.

\subsection{Accuracy of the absolute flux calibration}

Next, we check the absolute flux calibration of the present 9~$\micron$ point source catalogue.
The flux calibration of the limited number of sources yielding grade-S slit-less spectra was evaluated in the text (see Figure~\ref{flux_calibration_GradeS}).
In this subsection, we evaluate the overall absolute flux calibration as in Figure~\ref{flux_calibration_GradeS}, but using all sources  in the point-source catalogue.
Figure~\ref{flux_calibration_all} compares the fluxes in the present point-source catalogue with those in (a) the AKARI all-sky survey catalogue and (b) the WISE all-sky source catalogue.
The fluxes in the present 9~$\micron$ point-source catalogue match the all-sky survey at 9~$\micron$ within 1\%, and the WISE 12~$\micron$ within 5\%.
Hence, we reconfirm consistency of the 9~$\micron$ fluxes in the present study.

\subsection{Positional accuracy}

To verify the positional accuracy of the present 9~$\micron$ point-source catalogue, we compared the WCSs in the present catalogue and the WISE all-sky source catalogue (\citealt{Wright10}).
Figure~\ref{position}(a) presents the positional differences between the catalogues as a histogram.
After  fitting a Gaussian function to the histogram normalized by the circular ring area (Figure~\ref{position}(b)), the 1$\sigma$ standard deviation of the positional difference was determined as \timeform{0.31"}.
In principle, the positional accuracy in the present catalogue is not better than in the WISE catalogue (\timeform{0.15"} (1$\sigma$, 1 axis); \citealt{Wright10}).
On the other hand, the estimated accuracy is better than the pixel size of the reference image (\timeform{1.17"}).
Hence the position in the present 9~$\micron$ point-source catalogue is well determined.

\section{Formats of the spectroscopic and point-source catalogues}

The formats of the spectral and point-source catalogues are shown in Tables~\ref{spec} and \ref{psc}, respectively.

\section{Origin of variation among the spectra}

This appendix quantitatively examines differences among the spectra of the same objects (Figure~8), and validates the data reduction processes.
The spectroscopic toolkit considers three types of flux uncertainties; error in the spectral response curve, error of the positions on the reference image, and fluctuation of the sky signals.
The former two are systematic errors ($<$3\%) and the third one is random noise.
Therefore, the variations in the spectra for the same object are expected to be comparable to the random sky noise.

The variations in the spectra of the objects in Figure~\ref{stability} are quantified in Figure~\ref{stability_dependence}.
The S/N spectra of the ten objects (shown in the lower parts of Figure~\ref{stability}) were divided into seven wavelength ranges (5.5~$\micron$--12.5~$\micron$ in 1~$\micron$ increments), and examined the correlation between S/N and the flux in each wavelength range.
As shown in Figure~\ref{stability_dependence}, the S/N ratio improves with increasing flux, and is systematically lower in 11.5-12.5~$\micron$ range than in the other wavelength ranges.
We then fitted a power-law function ($y=ax^b$) to the data in each wavelength range, and estimated the 1$\sigma$ noise equivalent flux from the intercept of the best-fit function and the horizontal line at S/N=1.
We also estimated the 1$\sigma$-noise equivalent flux caused by fluctuations of the sky signals using the spectroscopic toolkit.
Using a function in the spectroscopic toolkit, we decomposed the noise components in the 16 spectra of J175421.97+664450.5, which was observed 16 times in the north ecliptic pole region (Figure~\ref{stability}), and calculated the average 1$\sigma$-noise equivalent flux spectrum.

Figure~\ref{sensitivity} compares the 1$\sigma$-noise equivalent fluxes estimated from the spectral variations and from the fluctuations of sky signals.
Although independently estimated, 1$\sigma$-noise equivalent fluxes of both estimates are generally comparable, indicating that the spectral variations of the same object in Figure~\ref{stability} arise from the fluctuating sky signals.
The 1$\sigma$-noise equivalent flux of the MIR-S slit-less spectroscopy estimated by 
\citet{Ohyama07} is also plotted in Figure~\ref{sensitivity}.
Considering that \citet{Ohyama07} estimated the 1$\sigma$-noise equivalent flux after applying smoothing over 2$\times$2~pixels to the dispersed images, we also confirmed that the 1$\sigma$-noise equivalent fluxes estimated in the present study and that estimated in \citet{Ohyama07} are roughly consistent.
Note that these analyzed data were collected from the ecliptic pole region, in which the sky signal from the zodiacal light is minimized.
Specifically, the intensity of the zodiacal light in the AKARI 9~$\micron$ band is $\sim$5~MJy/str at the ecliptic pole and $\sim$25~MJy/str in the ecliptic plane (\citealt{Kondo16}).
Therefore, the estimated 1$\sigma$-noise equivalent flux should be treated as the minimum value.
}


\clearpage

\clearpage

\clearpage

\clearpage

\clearpage

\clearpage
\begin{table}
\tbl{SIMBAD classification summary of the 604 objects that were spectrally analyzed in the present study.}{%
\begin{tabular}{cccc}
\begin{tabular}{cc}
\hline
	\# of objects & Classification \\
\hline  
    166 & *		\\
     15 & PM*	\\
     11 & G		\\
      8 & Y*O	\\
      6 & RG*	\\	
      6 & LP*	\\
      6 & IR	\\
      5 & X		\\
	5 & Sy2		\\
      5 & Sy1	\\
      5 & Rad	\\
      5 & LIN	\\
      4 & QSO	\\
      4 & Mi*	\\
      4 & C*	\\
      3 & V*	\\
      2 & Y*?	\\
      2 & TT*	\\
      2 & H2G	\\
      2 & EmG	\\
\hline 
\end{tabular} &
\begin{tabular}{cc}
\hline
	\# of objects & Classification \\
\hline  
      2 & AGN	\\
      2 & AB?	\\
      2 & **	\\
      1 & rG	\\
      1 & pr*	\\
      1 & pA*	\\
      1 & WU*	\\
      1 & WR*	\\
      1 & SN*	\\	
      1 & PoG	\\
      1 & PN	\\
      1 & HII	\\
      1 & Em*	\\
      1 & EB?	\\
      1 & DNe	\\
      1 & BD*	\\
      1 & ALS	\\
      1 & AB*	\\
      1 & *iC	\\
	316 & None  \\
\hline 
\end{tabular} \\
\end{tabular}}\label{classification}
\begin{tabnote}
Object classification in SIMBAD is shown in the following URL: http://simbad.u-strasbg.fr/simbad/sim-display?data=otypes
\end{tabnote}
\end{table}

\clearpage

\clearpage

\clearpage

\clearpage

\clearpage

\clearpage

\clearpage

\clearpage

\tabcolsep = 7 pt
\begin{landscape}
\begin{table}
\tbl{Format of the MIR-S slit-less spectroscopic catalogue}{%
\begin{tabular}{ccccccccccccccccc}
\hline
Work ID & Name & Grade & R.A. & Decl. & $F$ & $F_{\mathrm{err}}$ & $\mathrm{SF^{SG1}}$ & $\mathrm{SF_{err}^{SG1}}$ & $\mathrm{SF^{SG1}}$ & $\mathrm{SF_{err}^{SG1}}$ & Obs.Date & Obs.Time \\
 & & & (deg) & (deg) & (mJy) & (mJy) & & & & &(YYYY-MM-DD) & \\
(1) & (2) & (3) & (4) & (5) & (6) & (7) & (8) & (9) & (10) & (11) & (12) & (13) \\
\hline
3070002\_001\_008\_L& J000211.47+254550.7&  S&   0.54779&  25.76408& 2.516E+00& 1.51E-01& 0.72& 0.13& 0.64& 0.04& 2007-01-02&  0:20:49\\
3070002\_002\_010\_L& J000211.47+254550.7&  A&   0.54779&  25.76408& 2.778E+00& 1.67E-01&  NaN&  NaN&  NaN&  NaN& 2007-01-02&  2:00:06\\
3070002\_003\_009\_L& J000211.47+254550.7&  A&   0.54779&  25.76408& 2.801E+00& 1.68E-01&  NaN&  NaN&  NaN&  NaN& 2007-01-02&  3:39:24\\
3071014\_002\_017\_L& J000234.95+254950.0&  A&   0.64562&  25.83082& 1.790E+00& 1.07E-01&  NaN&  NaN&  NaN&  NaN& 2007-07-04& 11:05:53\\
3071014\_001\_012\_L& J000234.95+254950.0&  A&   0.64562&  25.83082& 1.825E+00& 1.09E-01&  NaN&  NaN&  NaN&  NaN& 2007-07-03&  3:36:11\\
\hline
\end{tabular}}\label{spec}
\begin{tabnote}
(1) Work ID ([observation ID]\_[sub-ID]\_[source number]\_[exposure]) (2) Object name in the 9~$\micron$ point source catalogue (3) Quality flag of the spectrum (4) Right Ascension (J2000) in the 9~$\micron$ point source catalogue (5) Declination (J2000) in the 9~$\micron$ point source catalogue (6) 9~$\micron$ flux measured with the aperture photometry for each reference image (7) Error of $F$ assuming the calibration error of 6\% (\citealt{Tanabe08}) (8) Absolute flux calibration factor applied for the toolkit-output SG1 spectrum (9) Error of $\mathrm{SF^{SG1}}$ (10) Absolute flux calibration factor applied for the toolkit-output SG2 spectrum (11) Error of $\mathrm{SF^{SG2}}$ (12) Observation date (13) Observation time
\end{tabnote}
\end{table}

\tabcolsep = 3 pt
\begin{table}
\tbl{Format of the 9~$\micron$ point source catalogue}{%
\begin{tabular}{ccccccccccccccc}
\hline
Name & R.A. & Decl. & $F_{\mathrm{aper}}$ & $F_{\mathrm{aper}}^{\mathrm{Ran}}$ & $F_{\mathrm{aper}}^{\mathrm{Sys}}$ & $F_{\mathrm{psf}}$ & $F_{\mathrm{psf}}^{\mathrm{Ran}}$ & $F_{\mathrm{psf}}^{\mathrm{Sys}}$ & $N_{\mathrm{det}}$ & $N_{1}^{\mathrm{source}}$ & $N_{2}^{\mathrm{source}}$ & 2MASS & WISE & Spec \\
 & (deg) & (deg) & ($\mu$Jy) & ($\mu$Jy) & ($\mu$Jy) & ($\mu$Jy) & ($\mu$Jy) & ($\mu$Jy) & & & & & & \\
(1) & (2) & (3) & (4) & (5) & (6) & (7) & (8) & (9) & (10) & (11) & (12) & (13) & (14) & (15) \\
\hline
J000156.64+255003.0&   0.48600&  25.83444& 2.525E+02& 4.06E+01& 1.51E+01& 3.850E+02& 4.78E+01& 2.30E+01&  1& 0&  0& 00015660+2550049& J000156.60+255004.8& 0\\
J000157.91+254906.5&   0.49128&  25.81848& 5.422E+02& 2.60E+01& 3.24E+01& 5.055E+02& 3.41E+01& 3.02E+01&  3& 0&  0& ---------------& J000157.88+254906.8& 0\\
J000157.92+254600.9&   0.49134&  25.76691& 1.569E+02& 3.92E+01& 9.37E+00& 5.620E+02& 6.45E+01& 3.36E+01&  1& 0&  1& ---------------& J------.---------.-& 0\\
J000158.45+254603.6&   0.49356&  25.76766& 6.346E+01& 3.77E+01& 3.79E+00& 6.167E+02& 6.97E+01& 3.68E+01&  1& 0&  1& ---------------& J------.---------.-& 0\\
J000201.91+254726.3&   0.50797&  25.79064& 4.760E+02& 2.54E+01& 2.84E+01& 5.498E+02& 3.66E+01& 3.28E+01&  3& 0&  0& 00020189+2547266& J000201.89+254726.5& 0\\
\hline
\end{tabular}}\label{psc}
\begin{tabnote}
(1) Source name (2) Right Ascension (J2000) (3) Declination (J2000) (4) Flux density measured with the  aperture photometry (5) Random error of $F_{\mathrm{aper}}$ (6) Systematic calibration error of $F_{\mathrm{aper}}$ (7) Flux density measured with the PSF photometry (8) Random error of $F_{\mathrm{psf}}$ (9) Systematic calibration error of $F_{\mathrm{psf}}$ (10) Number of detections (11) Number of nearby objects within a radius of 3$\arcsec$ (FWHM/2)  (12) Number of nearby objects within a radius of 20~pixels (\timeform{23.4"}) (13) Designation of the 2MASS catalogue (14) Designation of the WISE catalogue (15) MIR-S slitless spectrum is available (1) or not (0)
\end{tabnote}
\end{table}
\end{landscape}

\clearpage

\begin{figure}
 \begin{center}
  \includegraphics[width=0.7\textwidth]{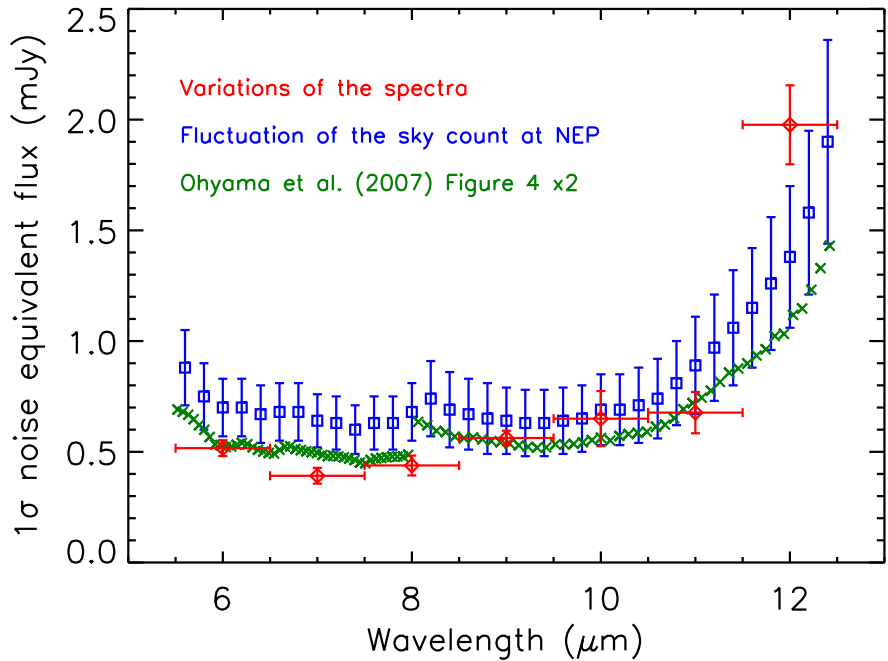} 
 \end{center}
\caption{1$\sigma$-noise equivalent fluxes as functions of wavelength, measured from (red) the spectral variations, (blue) fluctuation of the sky signals in the north ecliptic pole region, and (green) \citet{Ohyama07}. Since the noise equivalent flux estimated in \citet{Ohyama07} is reduced by a factor of two, the green data are doubled for a fair comparison.}\label{sensitivity}
\end{figure}

\clearpage

\bibliographystyle{aasjournal}

\end{document}